\documentclass[acmsmall,prologue,table,xcdraw,nonacm]{acmart}

\AtBeginDocument{%
  \providecommand\BibTeX{{%
    \normalfont B\kern-0.5em{\scshape i\kern-0.25em b}\kern-0.8em\TeX}}}

\setcopyright{acmlicensed}
\acmJournal{PACMHCI}
\acmYear{2021} 
\acmVolume{5} 
\acmNumber{CSCW2} 
\acmArticle{387} 
\acmMonth{10} 
\acmPrice{15.00}
\acmDOI{10.1145/3479531}

\usepackage{pifont}
\usepackage{graphicx}
\usepackage{adjustbox}
\usepackage{makecell}
\usepackage{longtable}
\usepackage{fancyvrb}

\begin{document}


\title{On the State of Reporting in Crowdsourcing Experiments and a Checklist to Aid Current Practices}
\titlenote{This is a post-peer-review, pre-copyedit version of an article accepted to the 24th ACM Conference on Computer-Supported Cooperative Work and Social Computing, CSCW 2021.}



\author{Jorge Ram\'irez}
\email{jorge.ramirezmedina@unitn.it}
\affiliation{%
  \institution{University of Trento}
  \city{Trento}
  \country{Italy}
}

\author{Burcu Sayin}
\email{burcu.sayin@unitn.it}
\affiliation{%
  \institution{University of Trento}
  \city{Trento}
  \country{Italy}
}

\author{Marcos Baez}
\email{marcos.baez@univ-lyon1.fr}
\affiliation{%
  \institution{LIRIS -- University of Claude Bernard Lyon 1}
  \city{Lyon}
  \country{France}
  }

\author{Fabio Casati}
\authornote{on leave from the University of Trento}
\email{fabio.casati@servicenow.com}
\affiliation{%
  \institution{Servicenow}
  \city{Santa Clara, CA}
  \country{USA}}

\author{Luca Cernuzzi}
\email{lcernuzz@uc.edu.py}
\affiliation{%
 \institution{Catholic University Nuestra Señora de la Asunci\'on}
 \city{Asunci\'on}
 \country{Paraguay}}

\author{Boualem Benatallah}
\email{boualem@cse.unsw.edu.au}
\affiliation{%
  \institution{University of New South Wales}
  \city{Sydney}
  \country{Australia}}

\author{Gianluca Demartini}
\email{g.demartini@uq.edu.au}
\affiliation{%
  \institution{University of Queensland}
  \city{Brisbane}
  \country{Australia}}

\renewcommand{\shortauthors}{Jorge Ram\'irez et al.}

\definecolor{ForestGreen}{rgb}{0.13, 0.55, 0.13}
\newcommand{\gd}[1]{\textcolor{ForestGreen}{GD: #1}}

\definecolor{royalblue(web)}{rgb}{0.25, 0.41, 0.88}
\newcommand{\jr}[1]{\textcolor{royalblue(web)}{JR: #1}}

\newcommand{\hl}[1]{\textcolor{black}{#1}}

\newcommand{\hll}[1]{\textcolor{black}{#1}}


\begin{abstract}
  Crowdsourcing is being increasingly adopted as a platform to run studies with human subjects. Running a crowdsourcing experiment involves several choices and strategies to successfully port an experimental design into an otherwise uncontrolled research environment, e.g., sampling crowd workers, mapping experimental conditions to micro-tasks, or ensure quality contributions. While several guidelines inform researchers in these choices, guidance of how and what to report from crowdsourcing experiments has been largely overlooked. If under-reported, implementation choices constitute variability sources that can affect the experiment's reproducibility and prevent a fair assessment of research outcomes. In this paper, we examine the current state of reporting of crowdsourcing experiments and offer guidance to address associated reporting issues. We start by identifying sensible implementation choices, relying on existing literature and interviews with experts, to then extensively analyze the reporting of \hl{$171$} crowdsourcing experiments. Informed by this process, we propose a checklist for reporting crowdsourcing experiments\footnote{The checklist can be found at \url{https://trentocrowdai.github.io/crowdsourcing-checklist/}}.
\end{abstract}



\maketitle

\section{Background \& Motivation}

Crowdsourcing platforms are being widely adopted as an environment to run experiments with human subjects \cite{DBLP:conf/chi/KitturCS08,DBLP:journals/sigkdd/MasonW09,Paolacci2010RunningEO,Crump2013}. 
\hll{Researchers are leveraging crowdsourcing} to test hypotheses, comparing different study methods, designs or populations, as well as to run studies aiming at observing user behavior. 
For example, crowdsourcing is helping researchers evaluate the impact of different interface designs on user performance, comprehension and understanding \cite{steichen2015supporting,dimara2017narratives,ramirez2019}, assess the difference in performance between users with different expertise, background and even mood levels \cite{wu2016novices,hube2019understanding,xu2019revealing}.   
These platforms (e.g., Amazon MTurk) give researchers easy access to a large and diverse population of participants, allowing them to scale experiments previously curbed to constrained laboratory settings.

\begin{figure}[h]
  \centering
  \includegraphics[width=\linewidth]{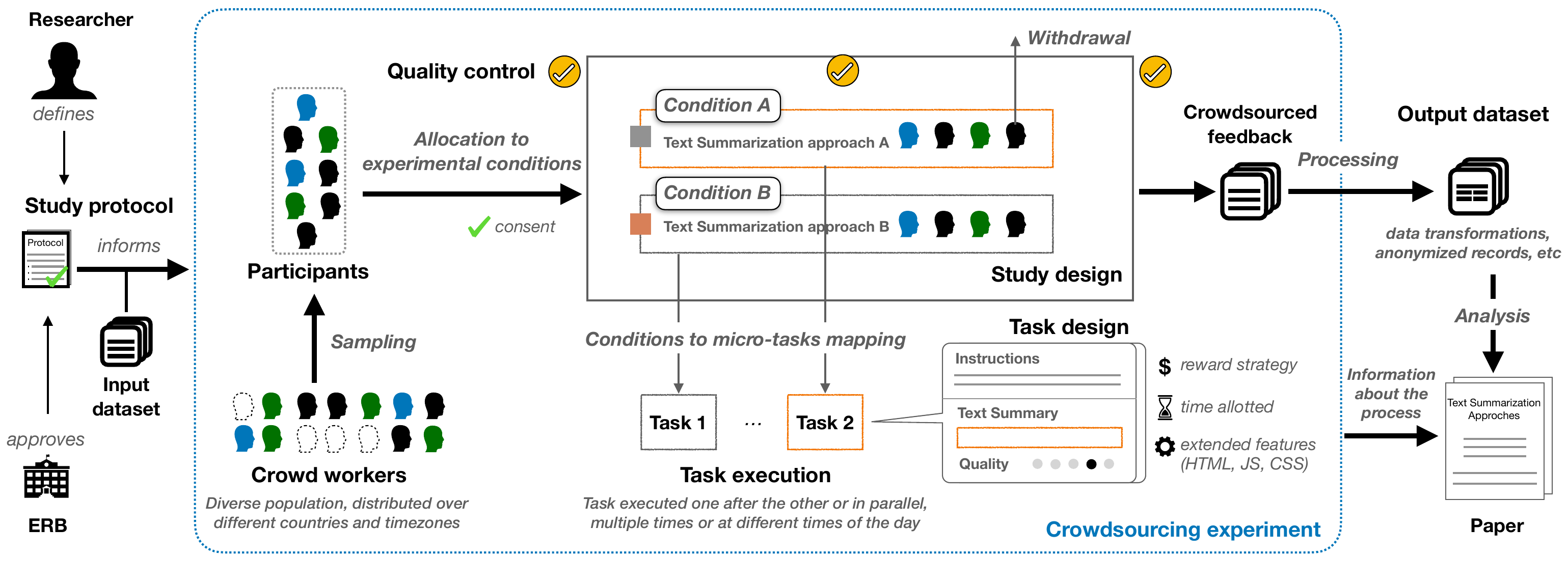}
  \caption{Mapping an experimental design to a crowdsourcing platform involves articulating many elements (none of which have a unique implementation). These elements constitute sources of variability if not properly reported.}
  \Description{Mapping an experimental design to a crowdsourcing platform involves articulating many elements (none of which have a unique implementation). These elements constitute sources of variability if not properly reported.}
  \label{fig:experiment-mapping}
\end{figure}

Researchers need to articulate many elements to successfully map and run an experiment in a crowdsourcing platform, as depicted in Figure \ref{fig:experiment-mapping}. 
The relevance of this is rooted in the need of incorporating more control and safeguards in an otherwise uncontrolled environment \cite{DBLP:conf/dagstuhl/GadirajuMNSEAF15}.
For example, an experiment testing the quality of two alternative approaches to text summarization would require researchers to define, among others, i) how to implement the two text summarization conditions as micro-tasks in the crowdsourcing platform (e.g., both conditions in the same task or in different tasks), ii) how to sample and allocate crowd workers to have representative, diverse (e.g., in culture, education or mother-tongue) and comparable groups assigned to both experimental conditions,\footnote{
\hl{While a perfect sampling and allocation strategy (of workers to conditions) may represent an ideal scenario detached from reality, we as researchers should ensure a reasonable balance in the underlying population to avoid confounds affecting the experimental outcome.}
} iii) quality control measures to avoid malicious or low-quality contributions from workers, and iv) task design and configuration, including the user interface, training examples, number of text summaries to show to each contributor, and the compensation for their participation. In doing so, researchers also need to ensure that their entire setup meets ethical standards for research with human subjects.

While guidelines and best practices have emerged to help researchers navigate the implementation choices and inherent challenges of running experiments in a crowdsourcing environment  \cite{DBLP:conf/icwsm/RogstadiusKKSLV11,Mason2012,Chandler2013}, little attention has been paid to how to \textit{report} on crowdsourcing experiments to facilitate assessment and reproducibility of the research.
If not properly reported, the elements mentioned above constitute sources of variability that can introduce confounds affecting the repeatability and reproducibility of the experiments, and preventing a fair assessment of the strength of the empirical evidence provided.

Reproducibility of experiments is essential in science \cite{DBLP:conf/chi/Wacharamanotham20}. The scrutiny of the research methods, by the academic community, and the development of standardized protocols and methods for communicating results are critical in the production of robust and repeatable experiments. Examples of this can be found in evidence-based medicine where systematic reviews follow a strict elaboration protocol \cite{Prisma}, or, more recently, in the machine learning community where authors are encouraged to follow pre-established checklists or datasheets to communicate their models and datasets effectively \cite{bender-friedman-2018-data,Gebru2019,DBLP:conf/fat/MitchellWZBVHSR19,DBLP:journals/ibmrd/ArnoldBHHMMNROP19,MLReproducibilityChecklist}.
Efforts in this regard have been branded under the definitions of \textit{repeatability}, \textit{replicability} and  \textit{reproducibility}, to denote attempts at obtaining 
similar results, by the same or different teams and experimental conditions, under acceptable margins of error \cite{plesser2018reproducibility}. Terminology notwithstanding, the importance of reporting experiments in sufficient detail has long been acknowledged as a fundamental aspect of research and of the scientific process.

The reporting of crowdsourcing experiments should be held to the same standards. Studies, under the umbrella of reproducibility and experimental bias, report on multiple aspects that affect the outcome of crowdsourcing experiments. These aspects include task design (e.g., instructions \cite{DBLP:conf/hcomp/WuQ17}, compensation \cite{DBLP:conf/www/HoSSV15}, task interface \cite{DBLP:conf/chi/SampathRI14}, and time \cite{maddalena2016crowdsourcing} among others), external factors such as the workers' environment \cite{DBLP:journals/imwut/GadirajuCGD17}, platforms \cite{Qarout2019PlatformRelatedFI}, and population \cite{DBLP:conf/wsdm/DifallahFI18} --- all of which serve as the foundation for running user studies.
In addition to the standard reporting for experimental research, it is paramount to identify and report the critical aspects and all possible knobs that take part in crowdsourcing experiments.
In this regard, existing literature has barely grappled with proposing guidelines that aid the reporting of crowdsourcing experiments, with existing guidelines limited to crowdsourced data collection in the social sciences \cite{Porter2020}.

In this paper we aim to fill this gap. 
We do so by identifying salient issues associated with the reporting of crowdsourcing experiments and propose solutions that can help address these issues. 
Specifically, we derive a \textit{taxonomy of attributes} associated with crowdsourcing experiments and turn this into a checklist for reporting. The checklist aims at facilitating the reporting of crowdsourcing experiments so that they can be repeated and so that a reader can assess if the experiment design matches the desired intent. 
The focus in this paper is on the reporting of  \textit{experiments} ran via crowdsourcing, that is,  studies aiming to answer research questions/hypotheses by following an experimental design, mapping the design to a crowdsourcing task and the features of the platform that supports it, and recruiting crowd workers as subjects.
%
%
\hl{This paper therefore excludes qualitative crowdsourcing studies (e.g., surveys) from its scope and, in general, other kinds of tasks that are not experiments (e.g., data labeling tasks are often not framed as experiments). However, the challenges in reporting on experiments are likely to be a superset of those or generic crowdsourcing tasks.}

\textbf{Contributions.} To aid the reporting of crowdsourcing experiments, we first need to understand the main factors that affect repeatability and that enable assessment of the quality of experiments by reviewers. 
We start by deriving the major design decisions of crowdsourcing experiments that play a role in crowdsourcing tasks and, therefore, on an experiment's outcomes.
We do so by following an iterative approach \hl{--- where we rely on literature reviews and interviews with experts ---} and derive a taxonomy of relevant attributes characterizing crowdsourcing studies.
Using this taxonomy, we analyze the state of reporting in crowdsourcing literature and identify aspects that are frequently communicated and those that tend to go under-reported. We leverage these observations to discuss potential pitfalls and threats to validity of experiments.
With feedback from experts, \hl{we then propose a checklist for crowdsourcing experiments to help researchers be more systematic in what they report.}
This checklist seeks to help experimenters describe their setup in a standardized format and readers to understand the used methodology and how it was implemented, serving as a tool that complements existing experimental research guidelines


\section{Related work}

\subsection*{Guidelines for reproducibility and reporting of scientific studies}

Reporting and reproducibility are at the heart of science.
Experiments allow researchers to manipulate a set of variables to test their influence into another group of variables of interest \cite{shadish2002experimental}. 
Guidelines for reporting scientific studies emerge from the observed variance in the methodological rigor associated with the studies. 
Indeed, the output of experimental research is only meaningful as long as it is reproducible \cite{DBLP:conf/www/Paritosh12}. This property guides the adopted methodology, as well as how this methodology and the results are communicated.

For example, in the study and synthesis of scientific results obtained via systematic literature reviews (SLRs), papers adhere to precise reporting guidelines that describes the (systematic) approach to investigating a problem as discussed in the literature.
The existence of study and study report protocols is what gives systematic reviews its methodological rigor, avoiding issues like a biased selection of clinical outcomes \cite{ChanEmpirical2014}. 
In this context, the PRISMA \cite{Prisma} guidelines propose a checklist to support the preparation and reporting of SLRs, making sure that such protocols exist and are reproducible.

In medicine, randomized control trials (RCTs) are the gold standard methodology to evaluate medical interventions. In this regard, guidelines like the CONSORT statement \cite{schulz2010consort} help authors properly report their RCTs and avoid potential issues resulting from the lack of methodological rigor (e.g., biased outcomes).

\subsection*{Guidelines in crowdsourcing contexts}

Research has shown how crowdsourcing could be leveraged in a wide range of tasks and domains (e.g., from labeling data for ML \cite{DBLP:conf/emnlp/SnowOJN08,DBLP:conf/cvpr/SorokinF08} to serving as a platform for experimental research \cite{DBLP:journals/sigkdd/MasonW09,Paolacci2010RunningEO,Crump2013}). 
Existing guidelines and best practices focus on how to run experiments in crowdsourcing environments successfully, proposing strategies to overcome common pitfalls found in crowdsourcing platforms \cite{DBLP:conf/chi/KitturCS08,DBLP:conf/icwsm/RogstadiusKKSLV11,Mason2012,Chandler2013}.
However, how to properly report crowdsourcing experiments has, to the best of our knowledge,  been somewhat overlooked. Our work complements experimental research in crowdsourcing by providing guidelines to aid researchers in reporting crowdsourcing studies.

Crowdsourcing acts as a surrogate to traditional participant samples, but with additional challenges that are not present in traditional experimental settings.
The quality of the contributions provided by crowd workers is a major concern due to the diversity in worker skills and the level of commitment crowd workers put into the task \cite{DBLP:conf/dagstuhl/GadirajuMNSEAF15}. Many quality control mechanisms have been proposed (e.g., see \cite{DBLP:journals/csur/DanielKCBA18} for a review) as well as studies analyzing performance as a function of internal and external factors, showing that intrinsic motivation could help in increasing the performance of workers \cite{DBLP:conf/icwsm/RogstadiusKKSLV11}.
Factors related to the design of the task could also contribute to obtaining subpar responses. Poorly written task instructions could misguide workers and result in low-quality work \cite{DBLP:conf/hcomp/WuQ17,DBLP:conf/cscw/KitturNBGSZLH13,DBLP:conf/ht/GadirajuYB17,DBLP:conf/naacl/LiuSBLLW16}. 
In contrast, enhanced interfaces may facilitate the job of crowd workers and aid these to improved performance \cite{DBLP:conf/chi/SampathRI14,Wilson2016WWW,ramirez2019,RamirezBMC2019}, as well as adequately limiting the time to judge can accelerate task completion without compromising the quality of the results \cite{maddalena2016crowdsourcing}.

Another challenge relates to how we operationalize an experimental design in a crowdsourcing platform. 
\hl{As opposed to laboratory settings, there is an inherent lack of control over the participants of crowdsourcing (and online) experiments that represents a concern to researchers, amplified by the absence of built-in support from crowdsourcing platforms \cite{DBLP:conf/chi/KitturCS08}.}
%
%
Random assignment, although simple in principle, is not straightforward to implement. Using multiple tasks to map different experimental conditions is a typical approach \cite{DBLP:conf/www/HoSSV15}; however, self-selection effects could arise due to participants preferring a subset of the conditions over others.
And ensuring that new workers arrive in the experiment is crucial to proper random assignment \cite{Mason2012,Chandler2013}, avoiding scenarios where workers participate multiple times in longitudinal studies or experiments that must run multiple times.
The characteristics of the platform should also be kept in mind.
The underlying demographics associated with the active workers \cite{DBLP:conf/wsdm/DifallahFI18} play an important role since differences in the conditions could be attributed to differences in the population of workers rather than the conditions themselves \cite{Qarout2019PlatformRelatedFI}.

Recent guidelines also safeguard the ethics behind crowdsourcing experiments.
Previous literature on humanizing crowd work pointed out relevant ethical issues present in common crowdsourcing practices (e.g., initially, crowdsourcing was seen as a ``cheap labor market'', as briefly reviewed in \cite{DBLP:conf/chi/BarbosaC19}). One aspect of this regards to accurate and fair task pricing \cite{Whiting2019FairWC}, which previous work has shown to impact the number of contributions produced by workers \cite{DBLP:journals/sigkdd/MasonW09}, the quality of their work \cite{DBLP:conf/www/HoSSV15}, and being a relevant aspect from an ethical perspective.
Another aspect concerns privacy, especially if some properties of the workers are being requested as part of the experiments \cite{Mason2012}.

As crucial as successfully conducting experiments in crowdsourcing environments, it is to make sure crowdsourcing experiments are reproducible \cite{DBLP:conf/www/Paritosh12,Qarout2019PlatformRelatedFI}.
In this regard, proper reporting of the methodology, operationalization, and results of crowdsourcing experiments plays a relevant role. 
Surprisingly, guidelines for reporting crowdsourcing experiments have received little attention even though underreporting is a serious concern in science, and crowdsourcing is no exception \cite{Buhrmester2018}.
Existing literature has identified how current studies fail to adequately report the methodology behind their crowdsourcing experiments, showing that most of the papers tend to omit information about worker qualifications, task design, rejection or validation criteria \cite{Porter2020,Ramirez2020DrecCSCW}.
These efforts propose templates for reporting studies, capturing essential aspects of crowdsourcing experiments.
However, these works are so far focused on specific use cases and platforms (data collection for social sciences in Amazon Mechanical Turk \cite{Porter2020}), or are still work-in-progress \cite{Ramirez2020DrecCSCW}.

Existing guidelines cover very well how to effectively leverage crowdsourcing in different contexts, but it has barely grappled with how to report crowdsourcing studies properly.
We aim to fill this gap by proposing guidelines for reporting the relevant aspects of crowdsourcing experiments. 
%
\hll{As a starting point, our work leverages on the preliminary taxonomy proposed in \cite{Ramirez2020DrecCSCW}. 
The final taxonomy we propose differs from DREC's in terms of validation and scope of the attributes.
The validation stems from a mixed approach involving a large-scale analysis of papers reporting crowdsourcing experiments, 171 articles from the literature, and feedback from experts in the field.
%
This approach helped us to scope down and refine the taxonomy to only attributes relevant to crowdsourcing experiments, leaving off attributes that are well-understood and covered in guidelines for experimental research 
\footnote{\hll{In the final taxonomy, the \textit{experimental design} dimension has 7 attributes vs. 13 in DREC. Here, the taxonomy omits attributes covered in guidelines for reporting study design and protocols (e.g., \cite{Gergle2014}). Likewise, the \textit{outcome} dimension, for example, omits the \textit{data analysis} attribute in DREC, as this is addressed by guidelines for reporting statistics (e.g., \cite{APA6thEDITION}).}}.
The final taxonomy also avoids narrow or too broad attributes (e.g., DREC's \textit{synchronous} and \textit{data analysis} attributes, respectively), as well as having attributes that can collapse into a single one (e.g.,  \textit{reputation} and \textit{environment} attributes are now part of the \textit{target population}).
In addition, our mixed approach allowed us to shape the final taxonomy into a checklist to aid how researchers report experiments in crowdsourcing.
}


\section{Data \& Methods} \label{sec:methods}

Our goals in this work are to 
i) understand the status of reporting on crowdsourcing experiments to assess which aspects are covered or neglected and the extent to which reporting is consistent across the literature, and 
ii) provide guidelines for reporting to assist in achieving consistency in current practices and facilitate reproducibility and assessment.



To achieve these goals, we draw inspiration from standardizing efforts in evidence-based medicine and software engineering for the reporting of systematic literature reviews \cite{Prisma,Kitchenham2007}. We follow a mixed approach that involves (1) deriving a taxonomy of relevant attributes characterizing crowdsourcing experiments, (2) leveraging this taxonomy to analyze \hll{171} papers published in major venues and get a picture of the current state of reporting, and (3) exploring potential alternatives of guidelines for reporting.





These steps were supported by  literature reviews and  interviews with experts in the field.
Specifically, literature reviews informed the first two steps, with a specific review for each step, and we describe them in detail in Sections \ref{sec:taxonomy} and \ref{sec:state}, respectively. 
The interviews with field experts provided a formative feedback along the entire process. They consisted of semi-structured interviews with researchers with
ample experience i) performing experiments to study crowdsourcing (i.e., crowdsourcing was the main area of research), or ii) leveraging crowdsourcing platforms to run user studies and experiments on different domains. Participants were recruited from the extended network of the authors, considering as eligibility criteria a research track involving crowdsourcing experiments and publishing in SIGCHI conferences.
\hl{Ten} experts agreed to participate (2F, 8M) including 1 Ph.D. student, 6 senior researchers from academia and industry, and 3 professors.

The interviews took place over Skype and Zoom between August and September 2020. Before the start of the sessions, participants were informed of the goal and scope the interview and provided their consent to participate and for the session to be recorded. The interview (see the protocol in Appendix \ref{sec:interview-protocol}) was organized into three parts that provided input to each of the three main steps in our methodology. The interviews were carried out independently by two researchers. All interviews were held in  and transcribed to English by the interviewer. Only the transcripts were accessed for the analyses. The analyses performed and how they inform our entire process are described in its proper context in the following sections. 




\section{A taxonomy of relevant attributes for crowdsourcing experiments} \label{sec:taxonomy}

This section introduces the taxonomy of relevant attributes that characterize different aspects of crowdsourcing experiments.
The attributes are grouped around six main dimensions denoting: the task requester, the experimental design used, the participants of the experiments (i.e., the crowd), the task design and configuration, the quality control mechanisms used to guard the quality of the results, and the outcome of the experiment.
%
The resulting taxonomy is summarized in Figure \ref{fig:taxonomy}. In the following we describe the methods and described in detail the final taxonomy, highlighting the literature support and expert opinions.

\subsection{Methods} \label{sec:methods-taxonomy}

We consider four sources to elicit the taxonomy: i) guidelines for research experiments in general and specific to crowdsourcing experiments, ii) features available in crowdsourcing platforms to support the deployment of experiments,  iii) scientific papers describing crowdsourcing experiments, and iv) interviews with experts.
With these sources we aim to convey in the taxonomy the landscape of elements taking part in crowdsourcing experiments: elements from experimental research, those inherent to crowdsourcing, and what features platforms offer.

\begin{figure}[h]
  \centering
  \includegraphics[width=\linewidth]{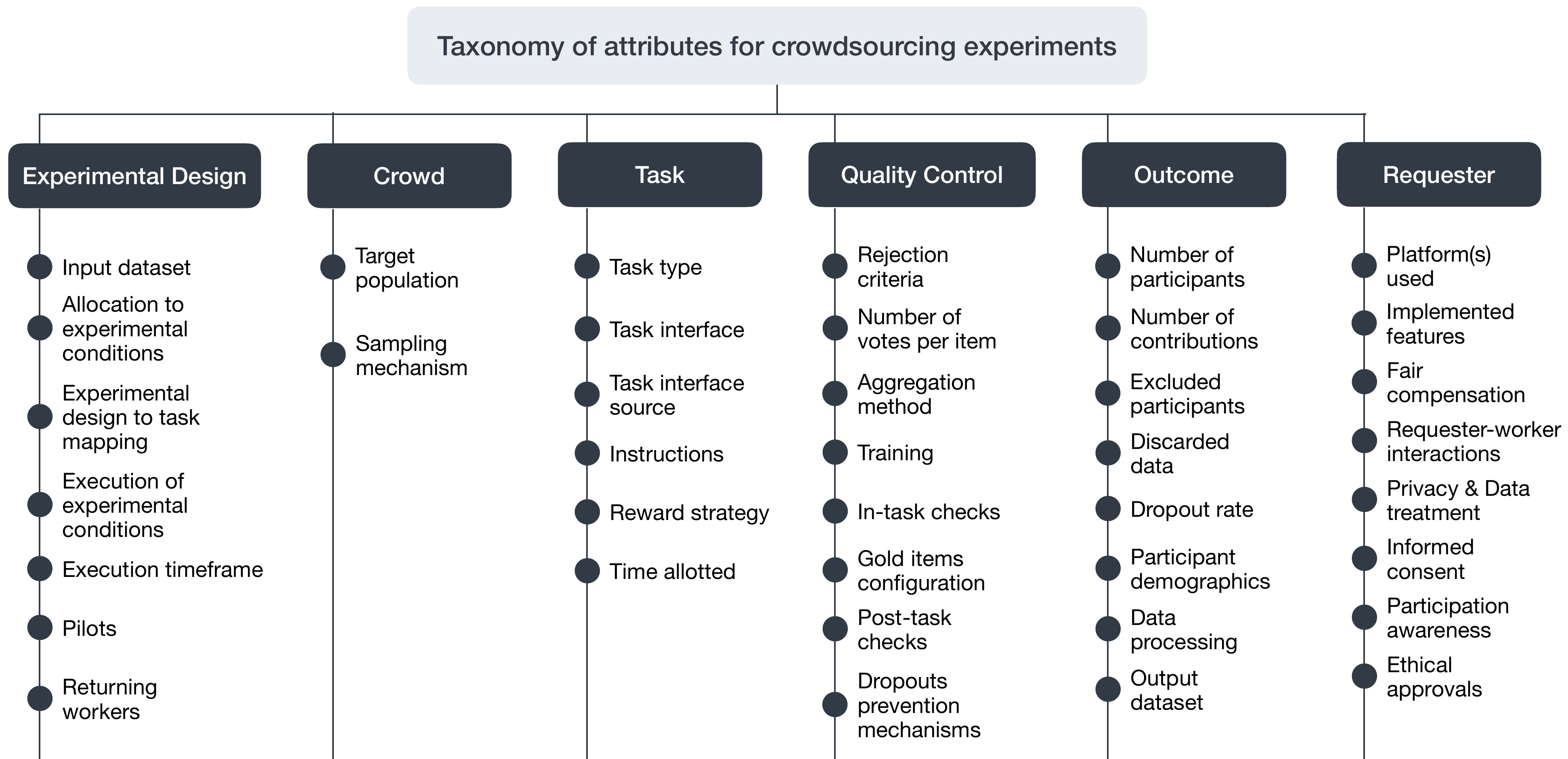}
  \caption{A taxonomy of relevant attributes characterizing experiments in crowdsourcing.}
  \Description{A taxonomy of relevant attributes characterizing experiments in crowdsourcing.}
  \label{fig:taxonomy}
\end{figure}

We started by identifying relevant attributes from existing guidelines. For this, we took a small seed of well-known guidelines for experimental design and crowdsourcing \hl{\cite{Gergle2014,Mahmood2015,Porter2020,Mason2012,DBLP:conf/dagstuhl/GadirajuMNSEAF15} and expanded it through snowballing and  keywords search using Google Scholar (\textit{``crowdsourcing + guidelines''}, \textit{``crowdsourcing + best practices''}, \textit{``crowdsourcing + recommendations''}, and \textit{``crowdsourcing + reporting + experiment''}) and screening the results based on title.}
This perspective was complemented with the analysis of practical task design attributes \hl{available in a example micro-task platform, Toloka~\footnote{\url{https://toloka.yandex.com/}}, which provides all common features for managing microtask crowdsourcing}.
\hl{The analysis of platforms' features allowed us to ground attributes to practical "knobs" that researchers need to consider to operationalize their experiments, and that often fall into assumptions (e.g., training steps, mapping of experimental design to concrete micro-tasks).}
Leveraging these sources, two researchers extracted an initial set of attributes \hl{(e.g., task interface, task instructions, compensation, crowd demographics, research design, random assignments, platform used, fair payments, among others)} in a spreadsheet to form an emerging list. The two researchers then jointly discussed and organized these attributes into six dimensions as depicted by the top-level entries in Figure \ref{fig:taxonomy}.

\hl{This organization of attributes required the two researchers to iteratively group the attributes identified in the seed of papers and selected crowdsourcing platform around common themes
(e.g., dimensions like \textit{pool of participants}, \textit{workers}, and \textit{study participants} were unified as the \textit{crowd} dimension). This was followed by an analysis of the initial set of attributes extracted, merging (whenever possible) equivalent attributes from different sources  (e.g., \textit{task template} and \textit{task UI} as \textit{task interface}).
}
\hl{Notice that during this process, the researchers defined the semantics and scope of each of the six dimensions, as well as that of the individual attributes.
We should stress, however, that our aim is on the comprehensiveness in terms of the attributes we identified and not the way we organized these into dimensions, which is a specific way of viewing things. 
For example, we include the demographics attribute under Outcome, as we define this dimension as the one capturing the \textit{results} of the different aspects of the crowdsourcing process (including the recruitment, application of quality control techniques, etc). Under different semantics, one might put demographics under the Crowd dimension, but we defined this dimension as the one capturing attributes and mechanisms to identify and sample participants from the Crowd.
}

This initial taxonomy was then refined by analyzing and piloting the extraction of relevant crowdsourcing experiment attributes from a list of research papers reporting crowdsourcing experiments. The process of identifying these papers is described in Section \ref{sec:methods-state}, as part of the analysis of the state of reporting. From this list, we took a random sample of 15 papers published in the last eight years, which we drafted incrementally until we reached saturation \cite{saunders2018saturation}. 
In this piloting phase, researchers took note of the applicability of the attributes for certain types of experiments, new attributes not initially considered, as well as attributes to be discarded or merged. These observations were discussed and addressed jointly by the two researchers.

We then further refined and validated this taxonomy with the input from crowdsourcing experts. To this end, we used the input collected in Part 1 and (some bits of) Part 2 of the semi-structured interviews with experts. The goal was to tap into their experience to identify, through different trigger questions, relevant attributes that we might have missed. 
We leveraged their input by inquiring about their experience running and designing crowdsourcing experiments (e.g., \textit{``What design choices you found to be more critical, possibly affecting experiment outcomes?"}), reporting or trying to replicate experiments (e.g., \textit{"What aspects you deem relevant and should be reported?"}), and their own experience reading or reviewing papers (e.g., \textit{"What aspects do you find to be typically under-reported or poorly reported?"}). In doing so, the interviewer would bring up only the top level dimensions (e.g., quality control or task design) if they were not discussed by the participant. We avoided providing any details about specific attributes so as not to bias the participants towards our taxonomy.
Then, Part 2 introduced a portion of the taxonomy (one or two dimensions) and asked participants to assess the relevance of the  attributes and to suggest any missing one.

\hl{
Transcripts were organized around the trigger questions in a document, where the interviewer highlighted excerpts touching on crowdsourcing attributes (e.g., \textit{``[...] depending on the \underline{design of the interface} you might get wide results.''}). These excerpts were then moved to a spreadsheet for analysis.
The interviewer then coded the excerpts with the associated attribute if covered by our taxonomy or flagged them for discussion otherwise.
The two researchers then discussed the coded excerpts and assessed whether i) the attribute was relevant to the scope of the taxonomy, ii) the scope and name of an attribute had be updated to cover a more general case, iii)  the scope of the attribute had to be limited to account for scenarios where an attribute is not applicable.
}

The inclusion criteria we used to refine the taxonomy considered three main reasons.
First, the taxonomy should focus on attributes directly associated with crowdsourcing, offloading general attributes to existing guidelines.\footnote{Initially, the taxonomy considered attributes such as hypotheses, independent, dependent, and control variables. In later iterations, we omitted these attributes since they are well understood and covered in guidelines for experimental research.}
Second, the focus should also be on practical and essential attributes for the reproducibility of experiments in crowdsourcing (e.g., instead of just describing the experimental conditions, authors should explain how these were mapped to tasks in crowdsourcing platforms).
And last, the attributes in the taxonomy should have a clear scope, avoiding ``unique cases'' or attributes that are too broad.\footnote{For example, the taxonomy considered whether an experiment was ``synchronous'' (experiments with multiple phases where one phase's output serves as input to the next \cite{Mason2012,DBLP:conf/dagstuhl/GadirajuMNSEAF15}). We ultimately replaced this by introducing more specific attributes describing how the experimental design maps to crowdsourcing tasks and how they are executed.}
The resulting taxonomy is the starting point to develop a checklist for reporting crowdsourcing experiments.

\subsection{Experimental Design}

The experimental design represents the building block for experiments of any kind, allowing to set the tone of the study and what level of conclusions can be derived from the experimental results.
Proper experimental research involves several elements, from the research questions to the study design and analysis.
These are well-understood aspects covered in experimental research guidelines and textbooks on research methods (e.g., \cite{Gergle2014, OlsonKnowingHCI}). 

The \textit{experimental design} section of the taxonomy we propose focuses instead on those attributes that are more closely related to crowdsourcing, and that allow experiments to be run in such platforms.
%
%
The limited support for running experiments in crowdsourcing platforms translates into many alternative strategies to map the experimental design to crowdsourcing tasks. 
Having multiple alternatives (and failing to accurately report these) can introduce confounds and ultimately affect the reproducibility of crowdsourcing experiments \cite{Qarout2019PlatformRelatedFI}. 
%
The \ding{108} \textit{experimental design to task mapping} attribute indicates how the experimental conditions (e.g., Condition A and Condition B) were mapped to crowdsourcing tasks. 
For example, as shown in Figure \ref{fig:experiment-mapping}, a between-subjects design could map each experimental condition to a different micro-task (Condition A $\rightarrow$ Task~2, Condition B  $\rightarrow$ Task~1) or randomize them within a single task (Conditions A,B  $\rightarrow$ Task 1).
Followed by this design choice, it is also relevant to define whether the experimental conditions were executed in parallel or sequentially (i.e., the \ding{108} \textit{execution of experimental conditions} entry in the taxonomy) as different execution strategies may result in experimental conditions leveraging different sets of active crowd workers (e.g., population samples with different underlying characteristics).

Another factor comprises participant's \ding{108} \textit{allocation to experimental conditions}. 
This attribute captures \textit{if} and also \textit{how} the randomization of the assignments was performed, considering the effect on the strength of the resulting experimental evidence. 
In crowdsourcing environments, where multiple micro-tasks may be running in parallel or sequentially to serve the overall experimental design, crowd workers may be able to engage in more than one micro-task, leading to what we refer to as \ding{108} \textit{returning workers}. 
Depending on the experimental design and how it was handled, this situation may be desirable or introduce unintended biases affecting the integrity of the experimental conditions and, ultimately, the experiment results \cite{Paolacci2010RunningEO,CrowdHub2019}.
Therefore, this entry in the taxonomy captures whether crowdsourcing experiments account for and communicate how returning workers were dealt with (or prevented in the first place).

The global scale of crowdsourcing platforms makes the active pool of workers to vary throughout the day \cite{DBLP:conf/wsdm/DifallahFI18}. Thus, failing to account for (and therefore report on) this aspect may add confounding factors to the experiment that hurt its reproducibility \cite{Qarout2019PlatformRelatedFI}. The taxonomy, therefore, needs to capture the \ding{108} \textit{\hl{execution timeframe}} of the experiment, by answering the simple question: over what timeframe was the experiment executed? (e.g., the experiment ran between January 1 and 10, every day at 2 PM).
The \ding{108} \textit{input dataset} fuels the task workers solve and is therefore crucial to indicate how this dataset was obtained and whether it is publicly available.
Lastly, the crowdsourcing literature advises to fine-tune a crowdsourcing experiment through multiple pilots \cite{DBLP:journals/jmlr/Vaughan17,DBLP:journals/sigkdd/MasonW09}, and it is intuitively relevant that authors report whether \ding{108} \textit{pilots} were performed before the main study.

\subsubsection*{\textbf{Expert opinion}} 
The interviews with experts organically touched on several of the experimental design attributes of crowdsourcing experiments, and while no new attributes emerged, their input allowed us to better scope and articulate the attributes. Their comments also provide a window into the challenges faced by experts in porting (and reporting on) experimental designs into crowdsourcing platforms.

Experts highlighted the importance of the experimental design in crowdsourcing experiments, indicating that ``\textit{the design is very critical}'', and at the same time acknowledging the challenges posed by the uncontrolled environment provided by crowdsourcing platforms. One participant illustrated this aspect while bringing up the importance of the allocation of crowd workers to experimental conditions
\begin{quote}
\textit{    ``In the ideal case, if you [had] a full control of the crowdsourcing platform, you would actually be able to do [a proper] randomization of crowd workers [to experimental conditions]. And being a requestor in a crowdsourcing platform such as AMT or Crowdflower, you don’t have this opportunity (...) Therefore, by not having these opportunities you kind of [perform] certain workarounds.''}
\end{quote}


The importance of experimental design to task mapping, execution dates, execution of experimental conditions and returning workers were all covered by the participants during the interview. One expert conveyed through an illustrative example how all these aspect are interconnected:

\begin{quote}
\textit{    Let's say you have two [conditions], like in an A/B test, and what differs in the two [associated] tasks is a design, literally a UI design. Then what you do is you run one task at 1pm today, and the other task tomorrow at 1pm as well. Well, time of the day is kind the same, but at the same time tomorrow [the platform would have a] slightly different population than today. It is very hard to control for some things. Maybe some participants that participate in your task today, will participate in your task tomorrow (recurrent workers), which definitely creates a certain bias and a ``feel of work" effect.} 
\end{quote}

Accessing input dataset and other more general aspects of experimental design not covered by the taxonomy (e.g., hypotheses, research questions) were also mentioned. One participant went as far as to say that even in ideal case where all information is available, even running the experiment under a different requester name could introduce bias \textit{``your requester name might be NASA (..) and people are just so excited to participate in studies run by NASA. It doesn’t mean that the effect is huge, but we know it [introduces] bias''. }

\subsection{Crowd}

The active pool of workers in crowdsourcing platforms constitutes the population of human subjects who can participate in crowdsourcing experiments.
%
The \ding{108} \textit{target population} aims to explicitly capture the eligibility criteria used to screen crowd workers and determine potential participants of the experiment. Conversely, this entry captures if no specific criteria were applied, thus implicitly using the characteristics of workers in the selected crowdsourcing platform as eligibility criteria (i.e., the entire crowdsourcing population is considered eligible). 
For example, studies may consider using specific demographic attributes, concrete environments in which workers perform their work (OS, web browser, mobile phones), a threshold to the task acceptance rate, or the number of tasks completed (typically provided by the platforms).

Once the target population is defined, the \ding{108} \textit{sampling mechanism} describes what strategies were used to recruit a diverse or representative set of participants from the target population.
Crowdsourcing environments give researchers more affordable access to a large and diverse pool of subjects with a wide range of demographic attributes, as opposed to laboratory settings where scale (and diversity) is constrained by time and available funds \cite{RAND2012ThePromiseOfAMT,DBLP:conf/dagstuhl/GadirajuMNSEAF15}.
This diversity plays an essential part in the external validity of an experiment, similarly to how the soundness of the methodology contributes to its internal validity (indeed, crowdsourcing experiments can be both internally and externally valid as their laboratory counterparts \cite{Horton2011}).
However, sampling strategies in crowdsourcing environments face additional challenges not found in traditional settings. For example, how the underlying demographics are in flux based on the active workers \cite{DBLP:conf/wsdm/DifallahFI18}; the lack of control over the subjects \cite{DBLP:conf/dagstuhl/GadirajuMNSEAF15}; and how easy it is for workers to quit, potentially causing non-random attrition and rendering the experimental conditions unbalanced \cite{RAND2012ThePromiseOfAMT}.

\subsubsection*{\textbf{Expert opinion}} 
The interviews with the experts addressed and enriched the two attributes associated to this dimension. 
Participants also emphasized how crucial (and challenging) it is to find the right workers who suit the needs of the study and can participate in it, and make sure a reasonable and representative sample is obtained from this target population.

Accordingly, both the worker profile making up the target population and the mechanisms used to sample participants were deemed relevant by participants, giving examples such as demographics (\textit{``I think is important to get to know some properties of the workers''}) and screening and qualification strategies applied to workers (\textit{``First is your recruiting requirement that includes the screening process: how you selected your participants, you need to describe that''}). 

Some participants went as far as to suggest that the mechanisms for inferring the properties of the target population, such as demographics and qualifications, are important and should be reported. For example, 


        
\begin{quote}
\textit{``Some tasks are designed in a so academic way. You are literally being asked `what do you see in the picture' [prompting the worker to annotate the picture], and then you have 25 questions about your age, income, race and something like that. I am a bit skeptical about this task design because first of all, it creates a huge bias. You know it's an academic study, you might really enjoy faking it, (..) or produce irrelevant results for example.
It might depend on [the] platform, in some you might get some data about the population.''}
\end{quote}

In relation to the above, another participant reflected on some uncertainties about using demographics, casting some light into the above behavior \textit{ ``it's not always clear if the gender of the participants is expected to be collected and reported, since not always [it] is relevant but it might be required by the reviewers''}.

The participants also provided specific examples of sampling strategies used in their experience, including sampling over time and geographical regions.


\subsection{Task}

The actual crowdsourcing tasks solved by workers can be regarded as the actual instruments or materials presented to the participants. The group of attributes presented here characterizes the tasks delivered as part of crowdsourcing experiments, considering organizational and operational details known to affect the results.

The nature and goal of the experiment shape the kind of tasks that are sent down to workers in crowdsourcing platforms, these aspects have been considered by previous research to propose a categorization of micro-tasks \cite{DBLP:conf/ht/GadirajuKD14}; we included in the taxonomy the \ding{108} \textit{task type} attribute to capture this information.

The \ding{108} \textit{task interface} in tandem with the \ding{108} \textit{instructions} concern the exact user interface and guidelines presented to workers. Poorly written instructions can misdirect workers and affect their performance in the experiments, resulting in subpar responses \cite{DBLP:conf/hcomp/WuQ17}. Besides, it can also reduce task intake and negatively affect how long the experiment takes to finish \cite{DBLP:conf/wsdm/HanRGSCMD19}. 
Variants of a task interface could unravel performance differences \cite{Mortensen2016crowd}. And similarly, current evidence suggests enriched interfaces may aid workers, allowing them to attain higher performance \cite{DBLP:conf/chi/SampathRI14,Wilson2016WWW,ramirez2019}.
The full disclosure of operational details such as task and instructions materials favors reproducible research. To this end, the taxonomy also captures the exact \ding{108} \textit{task interface source} (typically a combination of HTML, CSS, and JavaScript) uploaded to the crowdsourcing platform or related system (e.g., TurkServer \cite{DBLP:conf/hcomp/MaoCGPPZ12}).

The \ding{108} \textit{time allotted} workers to perform the task, as well as the \ding{108} \textit{reward strategy} used to motivate them, can also influence the progress of the experiment and resulting performance of workers.
In general, extrinsic factors such as proper monetary rewards can impact how much workers contribute \cite{Mason2012}. For effort-intensive tasks, in particular, adequate payments can motivate workers to produce results of higher quality \cite{DBLP:conf/www/HoSSV15}.
Low payments, however, may curb task intake and affect how fast experiments progress \cite{DBLP:conf/icwsm/RogstadiusKKSLV11,DBLP:conf/wsdm/HanRGSCMD19}.
Payment mechanisms are not limited to the amount being paid but also how. Different payment strategies could also lead to effects on how workers perform \cite{difallah2014scaling}.
Tasks can also include elements that target intrinsic motivational factors, for example, to aid how workers engage and commit \cite{DBLP:conf/lak/GadirajuD17} (although one could argue that intrinsic factors are effects we may want to reduce in an experimental setting unless it is the subject of study).
%
The time allotted workers to spend on the task naturally limits the associated cost of the experiment, besides studies in cost-effective crowdsourcing shed light on the impact of time on worker behavior and performance \cite{maddalena2016crowdsourcing,krishna2016embracing}.
We draw on these reasons to include in the taxonomy attributes that capture what mechanisms were employed to reward and motivate workers and whether time constraints where imposed.

\subsubsection*{\textbf{Expert opinion}} 
Aspects of task design such as the clarity of the instructions, the task interface and compensation were all suggested as critical design choices for the success of crowdsourcing experiments. Intuitively, participants  indicated that these aspects are relevant and must be reported.

Participants weighted in regarding different levels of information provided about the task interface, some mentioning that, for example, the reporting of the task interface should go beyond screenshots to include links to source files (\textit{``people report task design like a screenshot [...] It is maybe relevant to report the actual task design, like HTML, JavaScript, CSS files, so that people can reproduce it''}). Other experts, provided more nuanced opinions, e.g., 
\textit{``If there is a paper that talks about different treatments that are interface-related and there is no interface, it is a red flag. If there is no source code, it’s a yellow flag - for me at least.''}

In relation to the above, instructions and task descriptions were deemed critical. One participant illustrated how this information could completely shape the outcomes of an experiment, in the specific case of open-ended input: 
\begin{quote}
\textit{  ``The prompts and tips that we are giving before starting the task will definitely impact the final result. So, for example, if you give any useful examples of the feedback that crowd workers can generate, most probably, all of the workers will start imitating the same examples.'' } 
\end{quote}

Instructions, along with the payment and perceived effort (time) affect were reported as influencing not only the quality of the results but the decision to participate in the first place.

\subsection{Quality Control}

The varying skills, motivations, backgrounds, and behavior of online workers make quality control a major challenge in crowdsourcing \cite{DBLP:journals/csur/DanielKCBA18}. Therefore, quality control mechanisms are fundamental to any crowdsourcing task to safeguard the resulting quality of the contributions. 
An initial step concerning quality is to define what constitutes an invalid answer or contribution, and the \ding{108}~\textit{rejection criteria} aims to capture this information.  


Quality mechanisms can be set at different points in the crowdsourcing task. \ding{108} \textit{Training} protocols can be used to prepare workers for the job, which can have a positive impact on their resulting performance \cite{DBLP:conf/naacl/LiuSBLLW16}.
Another group of mechanisms can be categorized as \ding{108} \textit{in-task checks} or strategies embedded in the task to safeguard quality as contributions are collected. Whenever the nature of the task allows it, a fairly common practice is to embed gold items or attention checks to monitor the collected answers against the ground truth regularly. This simple technique can help to deal with malicious workers and avoid wasting collected contributions \cite{DBLP:conf/chi/KitturCS08}. 
The \ding{108} \textit{gold items configuration} attribute then captures how these items were selected (typically a subset of the input dataset), how frequently these gold items appear in the task, and what threshold was used to filter out workers underperforming on these items.
Other strategies involve imposing quality by design, such as adding feedback loops (where workers self-assess or receive external feedback) \cite{DBLP:conf/cscw/DowKKH12}, or making workers collaborate \cite{drapeau2016microtalk,schaekermann2018resolvable,Cicero2018}.

Another set of approaches are \ding{108} \textit{post-task checks} or mechanisms leveraged upon task completion, which can complement in-task checks and training protocols. 
Manual inspection (e.g., by an expert) can be used to review contributions from workers, limited by the scale of crowdsourcing and the cost associated with expert feedback \cite{DBLP:conf/cscw/DowKKH12}.
The assessment can also be computation-drive, for example, removing contributions that do not fall above a minimum time or agreement threshold \cite{DBLP:conf/websci/MarshallS13,DBLP:conf/cscw/HansenSCRG13}.

Depending on the type of task in the experiment, one may rely on redundancy, considering the \ding{108} \textit{number of votes per item} to be more than one, and leverage the same task to multiple workers to compensate potential noise. 
The \ding{108} \textit{aggregation method} comes in tandem with redundancy. A typical strategy is to use majority voting; however, more sophisticated and effective alternatives have been developed to derive the right answer even when the majority may be wrong (e.g., \cite{DawidSkene_Confusion,whitehill2009whose,Dong2013}).

Participants may drop out of experiments for different reasons and at different rates depending on the experimental conditions, introducing potential selection and attrition bias.
Different experimental treatments may be less appealing to workers and therefore result in non-random or selective attrition, introducing confounds in the experiment affecting the comparisons of the conditions \cite{Horton2011,RAND2012ThePromiseOfAMT}.
Copying with attrition, especially in online experiments, is instrumental to the study's internal validity, making sure dropouts stay roughly the same across conditions.
Therefore, the taxonomy captures what \ding{108} \textit{dropouts prevention mechanisms} were used to deal with the prevalence of task abandonment in crowdsourcing \cite{DBLP:conf/wsdm/HanRGSCMD19,kobren2015getting}.
For example, on Amazon Mechanical Turk, one can resort to using (neutral) qualification tasks to create a pool of potential participants and then send invitations to the actual tasks based on the assigned conditions \cite{DBLP:conf/www/HoSSV15}. 
Or, if one treatment incurs different effort levels than another, one may manipulate both treatments to include an effort-intensive activity and align how the difficulty of both tasks are perceived, regularizing the attrition rates \cite{RAND2012ThePromiseOfAMT}.

\subsubsection*{\textbf{Expert opinion}} 

Quality control was a hot topic during the interviews, cited as  \textit{``one of the biggest concerns in designing the tasks''}.
Participants considered quality control mechanisms as critical design choices that impact the results of experiments.  

The quality control mechanism suggested by the participants focused on the task flow, from strategies applied before (e.g., training or qualification assessment), during (e.g., gold items and attention checks) and after the completion of the task (e.g., filtering out contributions). In describing the control mechanisms applied to specific experimental settings, the participants highlighted that quality control mechanisms and when they are applied depend on the type of task. 
In some contexts, involving open-ended tasks such as content-creation, strategies such as gold items may not be applicable, requiring manual checks after task completion. 
\begin{quote}
    \textit{ ``In my experiments, the main problem - and that you would always have - is that there is no golden data for the answers that the crowd workers provide. When we are asking the crowd workers to provide open-ended text or feedback or anything, they can write anything they like. Here the quality control is very challenging and almost near to impossible to achieve.''}
\end{quote}

Indeed, a participant cited the exploration of automatic approaches as a \textit{``major area of research''} in his/her domain.

\subsection{Outcome}



The outcome dimension concerns details of the experimental results, more closely related to crowdsourcing, to aid their understanding, verification, and reproducibility.
The soundness of the data analysis process and how to articulate the findings aid to derive the right conclusions from experiments \cite{Gergle2014}. However, capturing these aspects are well beyond the scope of the taxonomy and refer them to available guidelines (e.g., \cite{StatisticsPrincipled1995,APA6thEDITION}).

The \ding{108} \textit{number of participants} and the \ding{108} \textit{number of contributions} collected (in total and across conditions) help to understand the scale of the experiment.
Also, the \ding{108} \textit{dropout rate} and \ding{108} \textit{demographics} quantify the level of attrition and diversity present in the experimental treatments, respectively.
Based on the rejection criteria, the experimenter may regard some of the contributions as invalid. The taxonomy identifies two related elements, the number of \ding{108} \textit{excluded participants}, perhaps malicious (or underperforming) workers, rendering a whole batch of answers from them as invalid. And \ding{108} \textit{discarded data}, i.e., specific contributions that were excluded before the data analysis.

While the previous attributes cover different quantities in the outcome, the \ding{108} \textit{data processing} captures any data manipulation step performed on the collected data. Intuitively, this information foster reproducible research alongside providing the \ding{108} \textit{output dataset} derived from the experiment (i.e., the raw or aggregated contributions from crowd workers).

\subsubsection*{\textbf{Expert opinion}} 
Interestingly, the comments from the participants would naturally flow more towards attributes that would allow researchers to repeat crowdsourcing experiments, and to a lesser extent the assessment of the outcomes.  
Among the few to touch on outcome attributes, two participants mentioned that the output dataset, as well as potential confounds like demographics, are relevant aspects of crowdsourcing experiments that should be reported (\textit{``it is important to report those aspects you did not control, such as demographic information''}).
%

\subsection{Requester}

The task requester is the person (or group of people) responsible for the design and execution of the experiment.
The organizational and operational details related to how the requester set up the experiment are also important.
The selected crowdsourcing \ding{108} \textit{platform(s)} constitutes the environment in which the experiment runs. It represents an essential element that guides how the requester operationalizes the study based on the features offered by the platform.
However, the available functionality tend to be limited for the requirements imposed by the experimental design \cite{DBLP:conf/chi/KitturCS08}, often requiring requesters to implement additional features to cover this lack of support. 
These \ding{108} \textit{implemented features} represent additional experimental instruments that also affect how feasible it is to replicate an experiment.

This limited support also translates into the available tools to facilitate interactions with workers (e.g., chat rooms and emails). 
\ding{108} \textit{Requester-worker interactions} could potentially impact how workers engage and perform in the tasks from the experiment \cite{DBLP:conf/cscw/DowKKH12}, so it is important to capture in the taxonomy if and how these interactions happen.
However, simple designs may not require complex interactions (or no interaction at all). 
For example, requesters may not need to interact with workers to study task design in the context of classification tasks, relying entirely on leveraging clear guidelines to articulate what they expect from workers \cite{ramirez2019,Mortensen2016crowd}.

Crowdsourcing environments define new legal grounds where policies may not be sufficiently defined (see \cite{felstiner2011working} for a review of the legal aspects around crowdsourcing).
The community is increasingly voicing the need for \ding{108} \textit{ethical approvals} for experiments run in crowdsourcing environments \cite{GraberIRB2013,DBLP:conf/cscw/MartinHOG14}.
\ding{108} \textit{Informed consent} tends to be often required for research with human subjects, as well as \ding{108} \textit{privacy and data treatment} statements for experiments that need to collect and store sensitive information.
\ding{108} \textit{Fair compensation} is also a relevant aspect from an ethical perspective. \hl{Computing fair wages indeed represents an active research area, considering that providing a minimum wage is not necessarily fair \cite{Whiting2019FairWC}.}
Initially, crowdsourcing platforms such as Amazon Mechanical Turk were deemed a ``marked for lemons'' \cite{IpeirotisLemons2010}, an environment restraining committed workers from earning at least a legal minimum wage due to the prevalence of less-committed or malicious workers.
Proper compensation can become a more frequent practice due to current features (like qualifications or badges) in crowdsourcing environments, advances in techniques to safeguard quality, plus underpayment issues being a reiterating topic addressed by recent literature \cite{DBLP:conf/cscw/KitturNBGSZLH13,DBLP:conf/chi/HaraAMSCB18,DBLP:conf/chi/BarbosaC19,Whiting2019FairWC}.
Additional context can be given to online workers, so they become aware they take part in an experiment. 
This \ding{108} \textit{participation awareness} is a relevant aspect because it is known to play a role in participant behavior \cite{McCambridge2012Awareness}.

\subsubsection*{\textbf{Expert opinion}} 
%
The operational context navigated by the requester was prompted by the participants in terms of the treatment of crowd workers, data management, and the technical environment. 

%
In terms of ethics and data management, participant stressed that workers should receive a fair compensation for their contributions and protected from potential harm, and that it should be clear that this is the case (\textit{``from the ethical perspective, [it should be reported] how much [crowd workers] were paid, whether they were exposed to certain content they were not naturally expected to see, like watching porn, cut bodies, and so on''}). Reflecting on the role of the different stakeholders in ensuring these practices, and whether ethical approvals should be required, some participant turned to practical recommendations:

\begin{quote}
    \textit{``I think [ethical] approval is a different thing. It is up to the institution where the researchers are from, but on the very generic case we just want to make sure that the crowd workers were told in advance [what they were exposed to]''}.
\end{quote}

%
The friction between observing the privacy of the crowd workers while making sure enough data was collected was also raised \textit{``I think is important to get to know some properties of the workers. But of course, it has implications in privacy''}.

As for the technical environment, participants expressed the importance of reporting from simple attributes, such as the actual crowdsourcing platform, to platform-specific features and complex configurations. One participant illustrated the latter for quality control configurations (\textit{``the whole quality control mechanism has a set of parameters that you have to set up, complex parameters that can lead to complex settings''}). But even for platform, it was argued that the role should be properly described (\textit{``You can say that F8/Append is kind of a hub for accessing other different platforms''}). 
Moreover, the case of F8 (now Append) serving as a hub to other platforms could be problematic if it is not reported properly, as indicated by the participant: 
\begin{quote}
    \textit{``Very few papers tell you, I ran my stuff on F8, but under the hood, F8 was forwarding the tasks to these 50 different platforms [...] These are like variables you add to your experiment and make the whole thing so complex''}.
\end{quote}



\section{Analyzing the state of reporting} \label{sec:state}

To study the state of reporting we surveyed \hl{171} crowdsourcing experiments from the literature, and assessed them on the basis of the relevant crowdsourcing experiment attributes from our taxonomy. 
We used the attributes to assess the \textit{completeness} and \textit{reporting style} used in communicating the attributes. By doing this, we aim to shed light on the level of reporting in current practices. In the following, we describe the methodology and results.

\subsection{Methods} \label{sec:methods-state}

%








The assessment considered the i) \textit{completeness}, referring to whether the attribute in question could be derived from the reported experiment, and ii) \textit{reporting style}, i.e.,  how the attribute was reported in the paper. We detail the process below. 

We started with a systematic search for scientific papers describing crowdsourcing experiments.
We iteratively refined a query for Elsevier's Scopus database to cover a wide range of crowdsourcing experiments in computer science and ensure the taxonomy fits well with a broad range of experiments.
The query consisted of keywords covering different usages or words associated with \textit{crowdsourcing} (e.g., ``crowd-sourcing'', ``micro-task'', ``human-computation'') and \textit{experiments} (e.g., ``experimental design'', ``study'', ``evaluation'', ``analysis'', ``intervention'').
We also limited the search space to papers published in major conferences (e.g., CSCW, CHI, IUI, UIST, HCOMP) between January 2013 and June 2020, \hl{excluding journal papers from the query to keep the search space focused}. 
We complemented this search by downloading from DBLP the list of papers in the proceedings of relevant conferences not indexed by Scopus.

The search identified a total of $670$ candidate papers. These papers were screened by two researchers to include papers describing experiments or user studies engaging crowd workers as subjects through a crowdsourcing platform. We excluded experiments leveraging on existing crowdsourced datasets, engaging workers in a role not related to the evaluation, experiments in laboratory settings, \hl{purely qualitative studies (e.g., surveys)} and experiments where crowd workers did not come from an open call to a crowdsourcing platform but rather from a more restricted environment. 
The screening started with a random subset of $50$ papers to calibrate the aforementioned eligibility criteria between the researchers, and then proceeded with the researchers screening independently the rest of the documents (the researchers agreed on 94\% of the decisions in the set of 50 papers, resolving disagreements by consensus). 
This process identified \hl{$172$} papers reporting crowdsourcing experiments. We refer readers to the Appendix \ref{sec:query-screening} for more details on the search and screening process, including the query and eligibility criteria.


%
The analysis then proceeded in three phases that aimed at assessing the reporting of experiments in terms of the taxonomy of relevant crowdsourcing experiment attributes. It started with a small sample of 15 papers, as explained in the above section, to iteratively refine the initial taxonomy and the assessment metrics. 
\hl{Second, to calibrate the interpretation of the taxonomy and the assessment metrics, a subset of 40 papers ($\sim$25\% of the total number of included papers) was analyzed by three researchers, resulting in an average agreement of 90\%. The researchers assessed all the taxonomy attributes for the same set of papers, considering as a match if they agreed on the completeness value (i.e., the presence of an attribute in a paper).
}
Finally, the rest of the documents were distributed equally among the researchers and assessed independently. Only one experiment was analyzed per research article, and in the case of papers reporting on multiple experiments, one experiment was randomly selected.   
\hl{Any doubts emerging during the independent tagging were flagged by the researchers for discussion and resolved by consensus. }

The analysis followed a meticulous procedure where the researchers leveraged on their experience (and the input from experts) to assess the completeness of the attributes of an experiment,  evaluating only those applicable to \hl{ i) the type of experiment reported in the paper, and ii) the objective of the study}. This approach was followed as it was clear from our pilots, interviews with experts and our own experience that not all attributes were relevant for each experiment. 
\hl{
With these criteria, the researchers ended up applying 26 of the 39 attributes in all cases, with the other 13 deemed not applicable (N/A) depending on the type and goal of the experiment.
We found that, on average, four attributes were N/A and that 80\% of the papers in our analysis had at most six N/A attributes.
Notice that we adjusted the total number of attributes on a per-paper basis, excluding the N/A attributes. Similarly, the percentages we report for each attribute in Figure \ref{fig:state} are also adjusted by removing N/A cases. We refer readers to Appendix \ref{sec:applicability-attr} for further details.
}

Taking the quality control attribute \textit{In-task checks} as a concrete example, the researchers first checked whether the paper explicitly mentioned any quality checks performed during the task (e.g., gold items, attention checks), and to the best of their abilities assessed whether they were indeed applicable for the particular type of experiment (e.g., the experiment studied cheating behavior where in-task checks would not make sense). Researchers then continued by checking if the details relevant to this quality control check could be derived from the description, even if the papers did not explicitly describe it (and in this case, it was counted as ``implicit reporting’’).
\hl{There are also some attributes with a ``directional association''. For example, we considered the interface and instructions as complete if the paper provided the source code (and, for example, omitted screenshots of these). Likewise, if \textit{informed consent} was explicitly reported, we considered as complete the \textit{participation awareness} attribute.}

\hl{
The assessment of the reporting covered not only the main content of the paper but also the appendix, supplementary materials and any source code or repository associated to the paper. To identify these additional sources, the researchers checked for links referenced or cited in the paper, as well as the official page of the paper in the publisher's website.
}
This procedure was followed to assess the completeness of all the attributes in the taxonomy.

\begin{figure}[h]
  \centering
  \includegraphics[width=\linewidth]{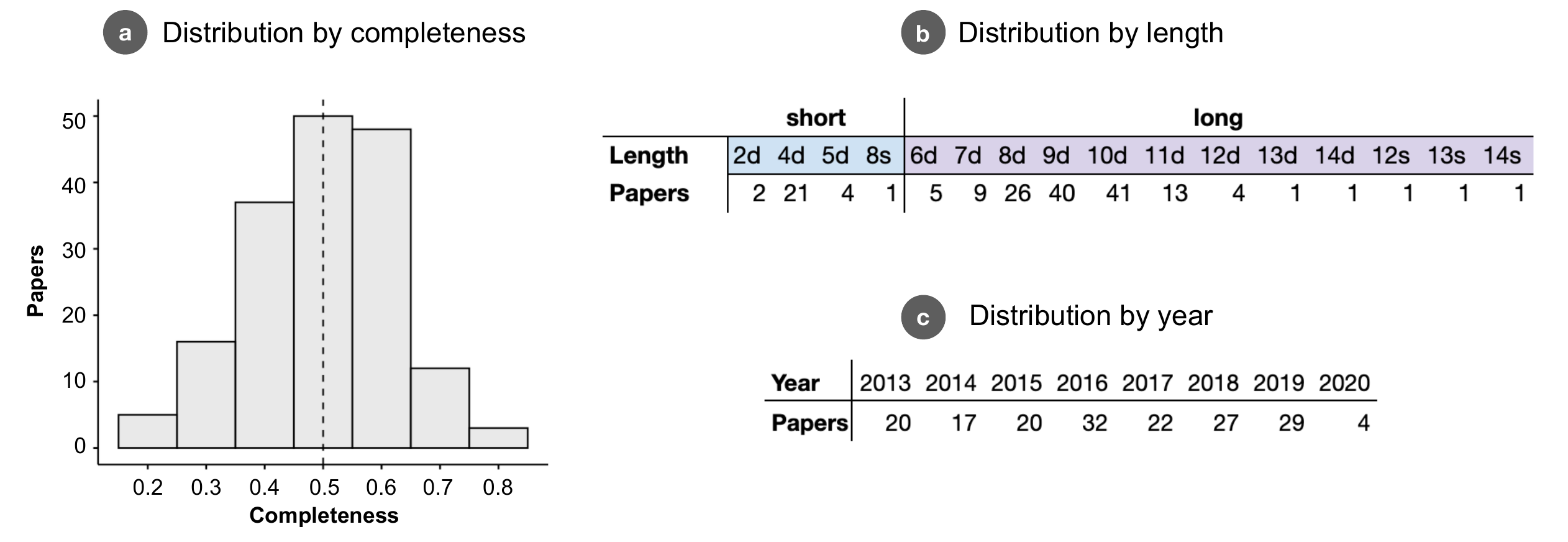}
  \caption{\hl{Some descriptive statistics of the 171 papers we analyzed. a) The distribution of papers by completeness, i.e., the proportion of attributes reported. The dashed line indicates the median. b) The distribution of papers based on their length (\textit{``d''}stands for double-column and \textit{``s''} for single-column format. c) The distribution of papers based on their year of publication.}}
  \Description{Some descriptive statistics of the 171 papers we analyzed. a) The distribution of papers by completeness, i.e., the proportion of attributes reported. The dashed line indicates the median. b) The distribution of papers based on their length (\textit{``d''} stands for double-column and \textit{``s''} for single-column format. c) The distribution of papers based on their year of publication.}
  \label{fig:descriptive}
\end{figure}

\subsection{Results}

\hl{We analyzed 171}\footnote{\hl{One paper was excluded from the analysis as the full-text was not available to the researchers.}} papers published in major venues to understand how (and to what extend) crowdsourcing experiments are reported. 
As mentioned previously, we used the attributes in the taxonomy to guide the analysis, assessing the completeness (e.g., \textit{is the attribute addressed?} And, \textit{is it explicitly reported?}), and the reporting style used in communicating the attributes (e.g., using screenshots, a figure, etc.).
By doing this, we seek to elucidate the level of reporting in current practices. 
\hl{We limited the analysis to 38 attributes, excluding the \textit{implemented features} attribute.\footnote{\hl{It is important to report on the features implemented to operationalize an experiment, but we found it difficult to reliably determine whether this attribute was applicable and being reported when analyzing the papers.}}}
%
%


\hl{Figure \ref{fig:descriptive} depicts some descriptive statistics of the papers we analyzed. The papers reported, on average, 49.9\% of the attributes (19 attributes), with 89 of the 171 articles reporting at least 50\% of the attributes. 
Of the papers analyzed, 89 of 171 were published between 2013 and 2016, and 82 of 171 were published between 2017 and June of 2020. 
Most of the documents (167 papers) were in double-column format and four in single-column format. The majority (143/171) are long articles (defined as those with \#pages $\ge$ 6 double-column, with 3 papers of 12+ pages long in single-column format), and 28 are short (\#pages $<$ 6 double-column, and one paper with 8 pages single-column). 
}

\hl{
We analyzed the level of reporting by year and length to contrast past and recent reporting efforts and differences based on paper length due to more pages available. 
We did not observe a clear pattern of recent papers addressing more attributes when compared to older ones, but we noticed an interesting trend of increasing reporting in longer documents. 
As for the attributes in our taxonomy, in 32/38, we noticed an increase in the percentage of papers reporting them, with relative differences of up to $3x$. However, the reporting was low, as the number of attributes covered by at least $50\%$ of the papers was $14$ for short and $16$ for long. 
Overall, the median completeness for short papers was $42.4\%$ and $51.5\%$ for long. 
Also, we did not observe any trend in the usage of supplementary materials to cover the lack of space, as only 15/171 papers provided supplementary material\footnote{\hl{The supplementary material included screenshots of the task (7/15), task source code (5/15), input dataset (8/15), output dataset (8/15), and other details related to data analysis (e.g., scripts, notebooks). Most of this material was provided as external links to a code repository (9/15) or document (3/15), two were in the appendix, and one was found via the publisher's digital library (video presentation).}}. Of these, three were short papers. 
A detailed breakdown of the results by year and length can be found at \url{https://tinyurl.com/ReportingState}.
}







%
%
%

\begin{figure}[h]
  \centering
  \includegraphics[width=\linewidth]{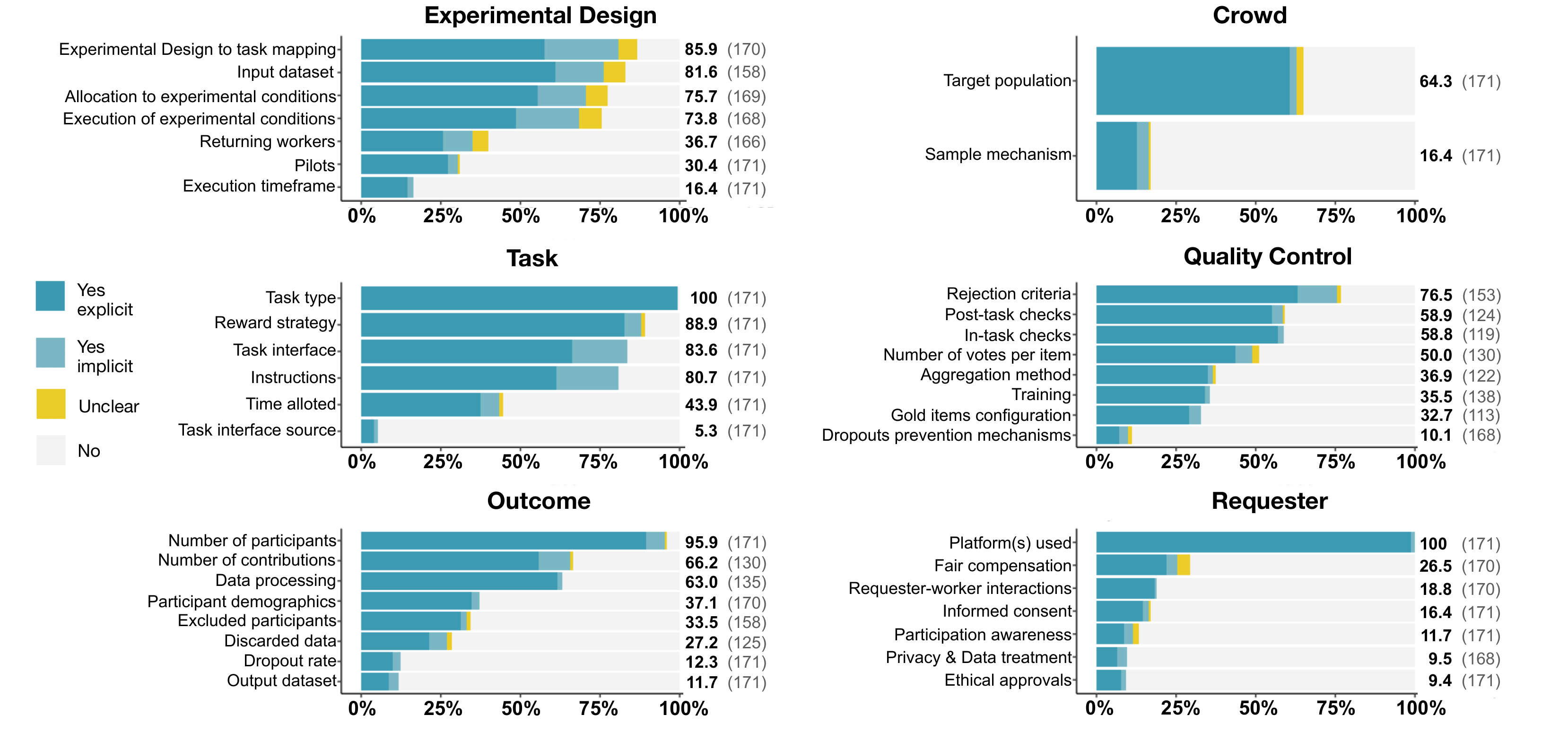}
  \caption{The state of reporting, based on the attributes in the taxonomy, for a \hl{set of 171} papers describing crowdsourcing experiments published in major venues. \hl{The number of papers to which the attribute applied is indicated in parenthesis, and in boldface, the percentages of papers reporting on it (explicitly and implicitly).}}
  \Description{The state of reporting, based on the attributes in the taxonomy, for a set of 171 papers describing crowdsourcing experiments published in major venues.}
  \label{fig:state}
\end{figure}

\hl{Figure \ref{fig:state} summarizes the analysis for each of the six dimensions in the taxonomy. In the following, we discuss the results for the attributes in each dimension, and when necessary, we will touch on interesting and relevant differences on the reporting level based on the year and length.}


\subsubsection*{\textbf{Experimental design}} Despite being the cornerstone of crowdsourcing experiments, the explicit reporting of experimental design attributes is for the most part shallow and unclear, as indicated by the low levels of explicit as well as high levels of unclear reporting -- the highest among the rest of the dimensions.
Overall, 4/7 attributes in the experimental design dimension -- those defining the design, mapping and execution of the experiment in the crowdsourcing platform -- are reported explicitly by \hl{52.4\% to 65.2\%} of the papers, representing the highest reported attributes for this dimension. When the experimental design is relatively simple and straightforward to map to a crowdsourcing environment (e.g., a within-subjects design mapping to a single crowdsourcing task), these four attributes can be derived with sufficient confidence, as indicated by the \hl{16.4\% to 24.7\%} of implicit reporting. But as soon as the design is more complex (e.g., a mixed design with multiple conditions and no clear hint on how these map to tasks), this is no longer the case, \hl{as indicated by the 0.6\% to 7.7\% of shallow and unclear reporting.} 


In terms of reporting style, the input dataset allowed for the richer set of approaches, compared to the rest of the attributes that were reported mostly in text.  A \hl{65.2\%} of the papers reported the actual input dataset used in the tasks given to workers, providing references or a link to an external site.  
However, some papers only indicated the process that was followed to prepare the input dataset and that would allow a researcher to construct a dataset of similar characteristics (which we coded as implicit, \hl{16.4\% of papers}). In \hl{two cases}, the provided link was no longer reachable, which brings a different issue: reporting links to supplementary materials that do not survive the ``test of time.''

Under-reported details include whether pilots were performed, if and how returning workers were controlled, and the experiments' execution timeframe. Pilots help tune the design of the experiment; however, it was only reported by \hl{27.5\%} of the papers. Returning workers, applicable depending on the design and how it is mapped, was reported by very few papers (only \hl{27.1\%}). Similarly, only \hl{14.6\%} of the papers reported the period in which the experiment run.
%
\hl{For long papers, all of the attributes show an increase in explicit reporting, with relative differences of up to $2x$. But, only 4/7 attributes are reported by at least 50\% of papers against 3/7 in short articles.}

\subsubsection*{\textbf{Crowd}} 
In general, \hl{62\%} of the papers explicitly reported properties characterizing the target population of workers suitable for the study, with \hl{2.3\%} of the papers implicitly suggesting what constitutes the target population \hl{(e.g., based on the goal of the study, we inferred that the task was open to everybody in the selected platform)}. Ultimately, this information was not clear in \hl{2.3\%} of the analyzed articles.
The exact mechanism used to sample a diverse and representative set of participants was reported in only \hl{12.9\%} of the papers (e.g., one possible strategy would be to sample systematically at different times of the day). \hl{In 3.5\% of the papers, this information was not explicitly addressed, but we derived by inference, which we coded as implicit (e.g., by providing the name of an external tool, by an auxiliary task to reach potential participants first).}
The above comes at a surprise, given the highly diverse and changing nature of crowd workers, and the potential influence of participants' characteristics and associated environment in task performance. In terms of reporting styles, these attributes are embedded in the paper as text. 
\hl{The target population was reported by $64.3\%$ of long papers vs. $50\%$ for short. And while the sampling mechanism was largely under-reported, it also shows an increase (14\% for long, and 7\% for short)}.


 \subsubsection*{\textbf{Task}} The task dimension is relatively well covered.
The type of task, the reward strategy, and how the task looks like (including the instructions) are the attributes reported by most of the papers. 
Of the papers analyzed, \hl{83.6\%} of them explicitly described the strategy used to reward workers for their contributions (almost all papers described monetary compensations as the reward strategy, except for \hl{three articles} resorting to volunteering).
The task interface was explicitly reported by \hl{66.1\% of the papers}.
Some articles (\hl{17.5\%}) did not explicitly include a screenshot of the task and instead described it partially in text (which we coded as implicit).
When reported, the interface is mostly depicted using screenshots and just a handful of papers \hl{(4.1\%) provided the actual source code of the interface, and in two cases the link provided was no longer available (coded as implicit)}.
But even relevant information such as the instructions is reported explicitly by \hl{61.4\%} of the papers. The instructions can be reconstructed from \hl{partial screenshots and textual descriptions in some cases (19.3\% of the papers)}. 
Explicit time constraints are reported in only \hl{38\%} of the papers.
%
\hl{In longer documents, we noticed an improvement in explicit reporting for details regarding the task interface (72\% for long, 35.7\% for short), instructions (66.4\% for long, 35.7\% for short), and the reward strategy (88.1\% for long, 60.7\% for short). The other attributes also show an increase, but the level of reporting is under 50\%.}




\subsubsection*{\textbf{Quality control}} Quality control mechanisms can take place before, during, and after the task. Despite their importance, the papers did not fully report the mechanisms employed.
\hl{Most reported attributes include the criteria used to reject contributions (rejection criteria, 64.1\%), the mechanisms used after the task to safeguard quality (post-task checks, 55.6\%), and in-task mechanisms for quality control (in-task checks, 57.1\%). Though we found cases where these attributes were addressed rather implicitly (12.4\% for rejection criteria, 3.3\% for post-task checks, and 1.7\% for in-task checks)\footnote{\hl{Examples of implicit reporting of rejection criteria would be when one could infer a paper actually accepted all contributions and filtered out on the data analysis part based on a metric like completion time, or a participant was not considered because they did not provide some demographic information.}}}.
\hl{Training sessions, when applicable, were reported by only 34.1\% of the papers. And details regarding the redundancy employed (number of votes per item) and aggregation method were reported by 44.6\% and 35.2\% of papers, respectively. In cases where gold items where used as quality checks, only 29.2\% of papers reported explicitly the configuration (gold items configuration).}
Dropout prevention mechanisms, closely related to engagement, were reported by very few papers \hl{(7.1\%)}. 
As for the reporting style, quality control attributes are described as text, with few cases relying on additional tables and figures to report information such as the number of votes per item (1 paper) and training sessions (2 papers).
\hl{Although, for the most part, still inadequate, there is an increase in explicit reporting in 6/8 attributes for long papers, though only two of these are reported by at least 50\% of papers (rejection criteria 64.6\%, and in-task checks 57.6\%).}


\subsubsection*{\textbf{Outcome}} 
The number of participants is reported in most cases (by \hl{90.1\%} of the papers), and by inference, the number of contributions, described explicitly in just \hl{56.2\%} of papers.
\hl{In some cases, 5.8\% of papers for number of participants and 10\% for number of contributions,  these details were not explicitly addressed, and we coded as implicit (e.g., the number of participants could be inferred based on the total number of contributions and the contributions per annotator).}
Data processing steps on top of the contributions from the crowd are reported by \hl{61.5\%} of the papers.
For the rest of the attributes, the papers do not paint a clear picture. For example, only \hl{21.6\% to 31.6\%} of the papers reported explicitly discarded data and excluded participants (due to quality or rejection criteria), and just \hl{9.9\%} described dropouts (workers leaving the experiment for different reasons). 
The output dataset and participant demographics were also poorly reported, with only \hl{8.8\% and 34.7\%} of the papers providing these details explicitly.
Papers reported the attributes mostly as text, accompanied by figures and tables for attributes such as the number of participants \hl{(21 papers) and contributions (11 papers)}, and participant demographics (5 papers).
\hl{For long documents, we noticed an increase in explicit reporting of 5/8 attributes, though, for the most part, reporting of these attributes is under 40\%, except for data processing (63.5\% for long, 50\% for short) and the number of contributions (57\% for long, 52.2\% for short)}.


\subsubsection*{\textbf{Requester}} Overall, the attributes in this dimension were vastly under-reported.
The papers reported the most basic information, the selected platform, but they poorly addressed the rest of the attributes (\hl{6/7 attributes reported explicitly by 6.5\% to 22.9\% of the papers}).
Most of the under-reported attributes relate to the ethics of the experiment. These attributes cover if compensation was fair (at least a minimum wage), whether workers gave their consent and were aware they took part in an experiment, if the study received ethical approvals and were in compliance with data privacy policies.
Regarding the reporting style, the papers described these attributes as text.
\hl{Interestingly, we noticed an improvement in recent years from an ethical perspective, when comparing papers from \textit{2013-2016} and \textit{2017-2020}. Reporting ethical approvals, fair compensation, and privacy and data treatment attributes increased. This insight aligns with current works addressing issues such as underpayments in crowdsourcing (see \cite{DBLP:conf/chi/BarbosaC19} for a brief overview). However, these improvements in reporting are still far from ideal.}


\subsection{Potential threats to validity}

The strength of the experimental evidence, as well as the ability of researchers to repeat and reproduce experiments, relies on the underlying methodology and how well this methodology and results are described. 
We have observed a reporting gap for the different dimensions in our taxonomy, which raises questions about experimental validity of crowdsourcing experiments reported in the literature. This section briefly discusses issues associated with under-reported attributes and how these connect to known biases from experimental research \cite{pannucci2010identifying} that affect the validity and integrity of scientific experiments.


\subsubsection*{\textbf{Sampling bias}} Researchers can introduce bias by inferring conclusions from a sample that is not representative of their target population. In crowdsourcing this can happen by not properly defining (or reporting on) a target population or because the mechanisms provided by the platform (or implemented by the researchers) fail to obtain a proper sample.    
As observed, the concrete target population and sampling mechanisms were not properly reported by \hl{38\% and 84.2\%} of the papers, respectively. The resulting demographics were omitted for \hl{62.9\%} of the experiments -- without weighing in the mechanisms to derive this information.
Even returning workers, under-reported by \hl{68.7\%} of the papers, may also bias the sample since these can hinder the goal of reaching a broader and more diverse set of workers.
The lack of proper reporting of these attributes not only affects the proper assessment of population samples, but it may also introduce practical challenges to replicability given the diverse characteristics of crowd worker populations.

\subsubsection*{\textbf{Selection bias}}
Bias can also occur due to the strategies chosen to assign participants to different conditions or cohorts. 
In crowdsourcing this could happen, for example, when the tasks associated to different conditions are executed under different operational settings (such as the ones mentioned in the following).    
Of the analyzed papers, \hl{83.6\%} did not report over what timeframe the experiment run, which is known to determine the characteristics of the active pool of workers \cite{DBLP:conf/wsdm/DifallahFI18}.
In this context, under-reporting how the conditions were executed, as in \hl{33.9\%} of the papers, can amplify this issue and render conditions with different sets of workers (e.g., conditions run sequentially with one capturing workers typically active during the morning and other conditions with night workers).
Task design attributes can also influence what type of workers are attracted to the tasks in the experiment. We observed \hl{19.3\%} of papers failing to report the instructions, \hl{12.3\%} the compensation, and \hl{57.3\%} the time constraints associated with the tasks. 
Reporting on these attributes is thus important, as they can reveal unintentional and intentional bias benefiting experimental conditions.

\subsubsection*{\textbf{Observation and response bias}} 
Also called the ``The Hawthorne Effect" \cite{sedgwick2015understanding}, observation bias arises when participants, being aware that they are taking part to a study, modify their behavior or contributions.
In a crowdsourcing experiment, this might be triggered when providing informed consents and acknowledging the scientific purpose of the task or, as made clear by the experts, in a more subtle way by the requester name, task design and even by the (fair) compensation.
In relation to this, participants might feel compelled to orient their responses towards what they believe the expected findings are, in what is called response bias. 
Only a few papers reported on participation awareness, and \hl{90.1\%} failed to report whether participants were informed of contributing to a study.

\subsubsection*{\textbf{Design bias}}
Researchers may fail to account for inherent biases present in experiments leading to what is called design bias. In crowdsourcing, in addition to other forms of experimental bias, there is a large body of literature on different forms of bias introduced by all aspects of task design and execution \cite{DBLP:conf/hcomp/WuQ17,DBLP:conf/www/HoSSV15,DBLP:conf/chi/SampathRI14,maddalena2016crowdsourcing,DBLP:journals/imwut/GadirajuCGD17,Qarout2019PlatformRelatedFI}. 
Our analysis shows that \hl{20.6\%} of the papers under-report how the experimental design was mapped to the selected crowdsourcing platform. 
Moreover, papers omitted information related to task design such as interface (\hl{16.4\%}), instructions (\hl{19.3\%}), and compensation (\hl{12.3\%}).
Even the implementation of desirable practices, such as running pilots to refine the experiments design, was reported explicitly in only \hl{27.5\%} of the papers.
This information gap can open the room to non-comparable conditions, especially when translating the experiment to other platforms with different features and workers.

\subsubsection*{\textbf{Measurement bias}} 
Bias can arise from measuring instruments of varying quality or errors in the data collection process. In crowdsourcing contexts, we can associate this bias to quality control. Overall, quality control attributes were under-reported, with \hl{41.9\%} of papers failing to report post-tasks checks, \hl{41.2\%} in-task checks, and \hl{64.5\%} did not describe training sessions.
Omitting these details may raise questions about the quality of the collected data, and allow for differences in quality when trying to rerun (or build upon) an experiment. For example, one could obtain better results by including training sessions.

\subsubsection*{\textbf{Ethical integrity}}
We observed that the vast majority of the papers failed to report on attributes such as fair compensation (\hl{77.6\%}), informed consent (\hl{84.2\%}), privacy \& data treatments (\hl{90.5\%}), and ethical approvals (\hl{90.6\%}). 
Not reporting on these attributes makes it difficult to assess whether experiments followed ethical guidelines and made sure workers were treated fairly and not exposed to any harm. This is clearly the ultimate goal. Understandably, processes for securing the ethics of experiments may vary from one institution to another, with some institutions enforcing ethical approvals on all studies while some put the burden on the researchers. What is clearly missing, and reflected on the interviews with experts and our own experience, are practical ethical guidelines that would allow researchers to make informed decisions about their crowdsourcing experiments, and proper support from crowdsourcing platforms to make sure these guidelines can be properly observed.

\section{A checklist for reporting crowdsourcing experiments} \label{sec:checklist}
The insights from the literature overview, expert interviews and the state or reporting, paint a daunting image for the reporting of crowdsourcing experiments, calling for the development of better guidelines and resources. In this section we take a small step in this direction, and describe the process that led to the development of a reporting checklist for crowdsourcing experiments. \hl{We stress that the goal of the type of support we explore here is to contribute to more a transparent reporting that can enable a better  assessment of the validity of experiments, as well as to repeatability. }

\subsection{Methods} \label{sec:methods-checklist}



The checklist is the result of a process that involved three main steps: i) internally growing the taxonomy into a reporting sheet that was used in the internal assessment and pilots; ii) obtaining feedback from experts on the generic reporting sheet and views on alternatives strategies for reporting, and iii) developing and refining the checklist. 

Starting from the taxonomy, two researchers defined each of the attributes and prepared questions that would require authors to specify if and how each attribute is reported in a paper. These  definitions were then used as the initial template for piloting the reporting of three papers of the authors, leading to the definition of the first ``sheet'' for reporting. The sheet contained the attributes in our taxonomy as rows along with their definitions. 
We filled out the sheet with excerpts from the papers explicitly addressing the attributes in our taxonomy. This step gave us a concrete example that we could use to discuss among the team members. This discussion led to a reframing of attributes and deciding what to include as part of the reporting. 
The discussion also led to brainstorming alternatives for the presentation, where we considered datasheets \cite{Gebru2019,DBLP:conf/fat/MitchellWZBVHSR19} or checklists \cite{Prisma,MLReproducibilityChecklist} as potential options for guidelines for reporting.


We focused Parts 2 and 3 of the interview with experts on assessing the definition of the attributes (as well as spotting missing ones) and collecting suggestions on how to present the guidelines for reporting.
Part 2 of the interview introduced participants with a portion of the final taxonomy (one or two dimensions). Participants (besides giving feedback on the relevance of the attributes and spot any missing one) were asked to read aloud attributes description and to assess and provide feedback on the framing.
Part 3 asked, \textit{``How do you think these aspects of crowdsourcing experiments could be framed into a tool for reporting?''}, providing examples such as datasheets or checklists.

The answers collected from these parts of the interview were used, first, to improve the definitions of the attributes and, second, to derive the final form of the guidelines: a checklist for reporting crowdsourcing experiments. 

\subsection{Proposed checklist}


    
    
    

The second and third parts of the interview with experts focused on 1) assessing the \textit{clarity} and \textit{completeness} of the attributes in the taxonomy, and 2) how we could exploit this taxonomy and turn it a tool for reporting. Based on the taxonomy and feedback we received in the interviews, we derived a checklist for reporting crowdsourcing experiments,
detailed in Appendix \ref{sec:checklist-table}.

The \textit{completeness} of the taxonomy makes sure the major ingredients of crowdsourcing experiments are well covered, while the \textit{clarity} concerns the interpretation and understanding of the attributes in the taxonomy. 
%
We found the taxonomy to be quite complete: the participants who identified missing elements (6/10) suggested attributes that were already present, but in other parts of the taxonomy they did not assess. 
One of the participants, \textit{P2}, suggested a ``when to stop'' attribute (not present in the taxonomy) for quality control, saying, \textit{``[...] when we are trying to collect quality data, we do not want to stop the task until we reach a threshold. Also, we might finish the task because we have already collected very good quality data"}. 
While we find this to be an excellent point, we argue that it is more related to data collection practices, for instance, to obtain data for ML, than for experiments in crowdsourcing. Therefore, we ultimately did not incorporate this aspect as an attribute in the taxonomy (besides, it could be covered by the existing post-task checks attribute).

\begin{figure}[h]
  \centering
  \includegraphics[width=\linewidth]{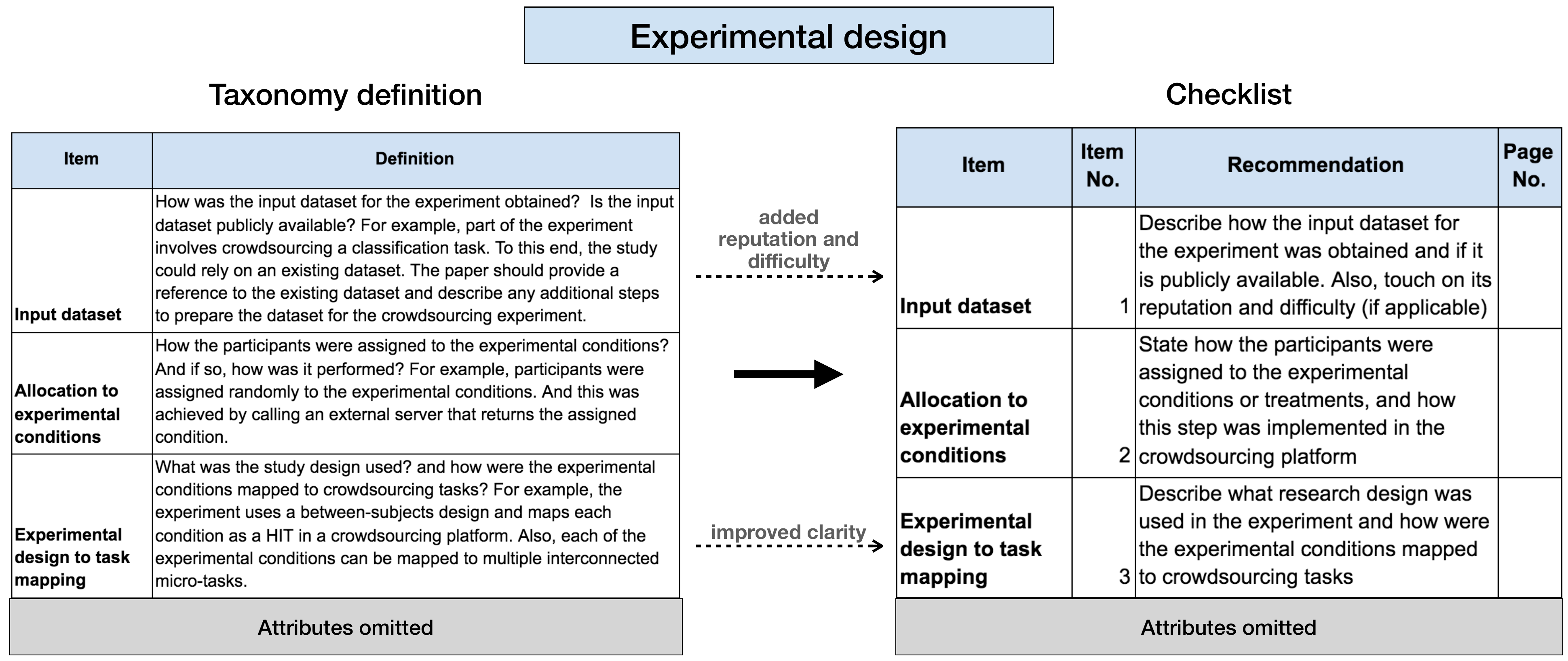}
  \caption{Example of updates we introduce to the attribute definitions in the taxonomy based on the feedback from the interviews (Experimental design dimension in this case), including the final framing as a checklist.}
  \Description{Example of updates we introduce to the attribute definitions in the taxonomy based on the feedback from the interviews (Experimental design dimension in this case), including the final framing as a checklist.}
  \label{fig:changes-exp}
\end{figure}

In general, the attributes were clear and relevant to the participants.
The suggestions of what attributes to add were used instead to improve the definition of the existing attributes, adding to the concrete feedback that focused on improving the definitions.
Figure \ref{fig:changes-exp} depicts some of the changes to the experimental design dimension and the final framing as a checklist.
Of the $39$ attributes in the taxonomy, $31$ were assessed by participants.\footnote{The requester dimension was not discussed with any of the participants. But aspects such as platform, ethics, and privacy were considered by some participants in the initial part of the interview.} 
%
Participants provided feedback, on average, \hl{to 10 attributes (between 7 and 14), with an attribute being assessed by 2 to 4 participants}.
%
We noticed that \hl{16 attributes were clear ``as is", 9 were clear, but received suggestions to improve them, 3 were clear only after some clarification from the interviewer, and only 3 were not clear}.\footnote{We coded an attribute as clear if it was clear to all the participants who assessed it. To code as ``clear with suggestions'', ``clear after clarification'', or ``not clear'', we only expected at least one feedback to fall in these categories, prioritizing the not clear option.}

As for how we could exploit the taxonomy, we asked participants, \textit{``How do you think these aspects could be framed into a tool for reporting?''}. 
We clarified that the framing naturally depends on the intended usage and shared two examples. 
\hll{The first example was the PRISMA checklist \cite{Prisma} that assesses the methodological rigor of systematic reviews.}
And the other was datasheets found in the ML community, which are self-contained structured summaries of a dataset creation pipeline or model performance \cite{Gebru2019,DBLP:conf/fat/MitchellWZBVHSR19}.
The participants favored the checklist format \hl{(5/10)} over the datasheet alternative \hl{(1/10)}, while the rest did not explicitly mention a checklist but suggesting it based on their answer.

\subsection{Intended usage and adoption}


\hl{The intended usage of the taxonomy, how it is framed as a tool for reporting, and the adoption by the research community go hand in hand.}

\hl{A checklist format gives paper authors more flexibility to describe the different aspects of the experiments in the paper's main content and supplementary materials (weighing in typical page limits), indicating where an attribute is being described. We aim for the checklist to serve as a resource that guides researchers in what they report, helping them be thorough and systematic in communicating the details of their crowdsourcing experiments (i.e., serving as a reminder,  \textit{``when we are very deep into our data analysis part we forget the basic stuff that should be reported''}).}
Unlike a datasheet, a checklist is not self-contained, which was indicated by one of the participants, \textit{``To me, it is easier to see it as a report, with all the information on the same page [...]''}. 
While a self-contained summary as a supplementary material could be more convenient to readers, it demands more effort from authors since the main ingredients of the experiment would still need to be described in the paper, at least at a high level.
Guided by the feedback from the experts, we ultimately opted for the checklist format, considering that it helps authors report their experiments, while avoiding additional efforts, and readers to navigate the details of a crowdsourcing experiment.
%



\hl{
During the interviews, the participants raised challenges associated with the adoption of a checklist --- \textit{``it all comes down to motivation''}.
%
One of this challenges was associated with the research community and the current practices around crowdsourcing experiments. As explained by a participant:
\textit{``you can probably check 30\% of those boxes and the paper will be published [...] no major motivations in the academic world other than your paper being published''}.
Adopting a checklist would go in tandem with evolving current community practices for assessing papers reporting on crowdsourcing experiments (as seen, for example, in the ML community, where reproducibility checklists are part of the submission/reviewing process\footnote{\url{https://neurips.cc/Conferences/2021/PaperInformation/PaperChecklist}}). However, we hope that bringing awareness about the potential issues of not standardizing practices can push the community in the right direction.
In response to this challenge, the experts highlighted the importance of growing adoption organically by making people aware of the benefits and lowering the barriers to adoption. As commented by the participants: \textit{``convince the people that these are the important things that you should follow''}, and \textit{``make researchers' life easier''}.
}

\hl{
We foresee promising avenues of future work that could address these challenges and facilitate adoption by the research community. In this work, however, we limited ourselves to starting the conversation towards reproducible crowdsourcing experiments.
For instance, convincing people to adopt the checklist could be addressed by providing empirical evidence on how using the checklist aids reproducible results. Or making its usage so trivial that researchers just adopt it. For example, \textit{``As a way to make their lives easier, you say, hey, tell me the platform, tell me the task ID, provide credentials, click ENTER, and a GitHub repository is created with all this information''}.
}


\section{Discussion}



\hl{
The current state of reporting in crowdsourcing research still misses providing details beyond basic attributes associated with task design, quality control, requester, and experiment design and outcome.
According to our analysis, at least 70\% of papers report the selected platform, how the experimental design maps and executes, reward strategy, task interface, instructions, rejection criteria, and the number of participants. However, if we consider only explicit reporting, this list narrows to reward strategy, the number of participants, and the selected platform.
While these attributes are relatively well-covered, either explicitly or implicitly, most tend to be under-reported by at least 50\% of the papers. Among the six dimensions in our taxonomy, the requester --- with attributes covering the ethics of experiments --- was among the least reported by the papers. These issues open the room for potential threats to validity associated with missing details regarding the experimental design and its operationalization.
}

\hl{
Under-reporting poses the interesting question of why the attributes are poorly reported in the first place and how we can overcome this situation.
Our analysis and feedback from experts attribute this issue to the limited guidance and awareness on what needs to be reported. 
By providing a checklist and depicting where the research community stands in terms of reporting practices, we expect our work to stimulate additional efforts to move the transparency agenda forward.
It is clear from our study that improved transparency in reporting is a shared responsibility among the different stakeholders in the crowdsourcing ecosystem.
}

\hl{
\textbf{\textit{Researchers}} may be asked to agree to a code of conduct or follow guidelines, like our checklist, to improve the current level of reporting. 
As a \textbf{\textit{research community}}, we need to develop guidelines and best practices (e.g., on how we set up and report experiments) to increase transparency, strength, and reproducibility of crowdsourcing experiments. And such guidelines should also emphasize the ethics and fairness behind experiments.
In turn, \textbf{\textit{venues}} need to enforce and adopt higher reporting standards, mirroring the initiatives taking place in other communities.
}

\hl{
\textbf{\textit{Platforms}} can benefit from the insights and design recommendations emerging from our work and address operational barriers to reporting.
Platform providers may aim for solutions that are ``experiment-aware'' (e.g., by offering features to treat experiments as first-class citizens).
Indeed, platforms can play a major role in both i) helping task requesters to design experiments that are consistent with accepted ethical guidelines (from informed consent to minimum wage) and ii) helping to generate reports that facilitate publication of relevant experiment information to aid reproducibility.
}

\bigskip

\noindent \hll{
\textbf{Limitations.} It is important to note that the assessment of crowdsourcing experiments in this work is limited to what is reported, which might be an incomplete picture of the design and operationalization of the experiments --- but that is what is eventually available to the community to build upon and replicate.
%
While a systematic process was followed to cover as much research landscape as possible, the search can not be considered an exhaustive account of crowdsourcing experiments in the literature. Yet, the 670 conference papers screened and the 171 analyzed in detail provide a representative sample of the current state of affairs.
Also, it is worth noting that we did not manage to cover the four stakeholders we mentioned before, as we did not interview people (e.g., managers and crowd workers) from major crowdsourcing platforms. However, the participants we interviewed possess ample experience designing and executing crowdsourcing experiments in these platforms, with research output published in major SIGCHI conferences; they helped provide thoughtful design and operational angles to the attributes in the taxonomy and the final checklist.
}

\section{Conclusion}

The takeaway message is that the characteristics of the crowdsourcing environment pose several challenges to the design and execution of experiments, making them subject to more complex sources of bias. 
Therefore, it is of paramount importance to encourage further transparency to aim for higher standards of evidence and reproducibility of results. As we observed, however, we still lack in this regard
\hll{--- under-reporting leaves} room for potential threats to validity. This invites a reflection on how valid are current crowdsourcing experiments, how reproducible they are, and whether they comply with ethical standards that ensure fair treatment of crowd workers and their data.

This work pushes crowdsourcing research towards standardized reporting of crowdsourcing experiments. Current guidelines focus on effective task design and practical recommendations for running experiments. But these guidelines somewhat overlook advising researchers on what and how to report the details underlying crowdsourcing experiments. 
This paper introduced a taxonomy of relevant ingredients characterizing crowdsourcing experiments and used this taxonomy to analyze the state of reporting of a sample of articles published in top venues. This process allowed us to identify gaps in current reporting practices. To help address these issues, we leveraged the resulting taxonomy and feedback from experts to propose a checklist for reporting crowdsourcing studies.

Promising future work avenues include evaluating if and how the checklist aid reproducible research, exploring methods to automatically derive crowdsourcing experiment reports from existing crowd platforms, and  developing a system with crowdsourcing experiments as first-class citizens.

\begin{acks}
The authors thank our interview participants and anonymous reviewers for their helpful comments and feedback.
\end{acks}


\received{October 2020 }
\received[revised]{April 2021}
\received[accepted]{July 2021}

\bibliographystyle{ACM-Reference-Format}
\bibliography{shorter-references,bibliography}


\appendix
\section{Checklist for crowdsourcing experiments} \label{sec:checklist-table}

In this section, we introduce the checklist, depicted in Table \ref{tab:checklist}. The checklist should be filled out per experiment, in case the paper reports on multiple studies involving the crowd as subjects. Besides, suppose an experiment uses different (potentially interconnected) micro-tasks. In that case, the Task and Quality control sections should be reported per task (or at least the Task section in case the quality control mechanisms are the same for all tasks).

\begin{longtable}{| p{.25\linewidth} | p{.05\linewidth} | p{.50\linewidth} | p{.05\linewidth} |}
\caption{Checklist for reporting crowdsourcing experiments}~\label{tab:checklist}\\
\hline
\rowcolor[HTML]{CFE2F3} 
\multicolumn{1}{|c|}{\cellcolor[HTML]{CFE2F3}\textbf{Item}} & \multicolumn{1}{c|}{\cellcolor[HTML]{CFE2F3}\textbf{\begin{tabular}[c]{@{}c@{}}Item \\ N.\end{tabular}}} & \multicolumn{1}{c|}{\cellcolor[HTML]{CFE2F3}\textbf{Recommendation}} & \multicolumn{1}{c|}{\cellcolor[HTML]{CFE2F3}\textbf{\begin{tabular}[c]{@{}c@{}}Page \\ N.\end{tabular}}} \\ \hline
\multicolumn{4}{|l|}{} \\ \hline
\rowcolor[HTML]{CFE2F3} 
\multicolumn{4}{|c|}{\cellcolor[HTML]{CFE2F3}\textbf{Experimental design}} \\ \hline
\textbf{Input dataset} & 1 & Describe how the input dataset for the experiment was obtained and if it is publicly available. Also, touch on its reputation and difficulty (if applicable) &  \\ \hline
\textbf{Allocation to experimental conditions} & 2 & State how the participants were assigned to the experimental conditions or treatments, and how this step was implemented in the crowdsourcing platform &  \\ \hline
\textbf{Experimental design to task mapping} & 3 & Describe what research design was used in the experiment and how were the experimental conditions mapped to crowdsourcing tasks &  \\ \hline
\textbf{Execution of experimental conditions} & 4 & Report how the crowdsourcing tasks, representing the experimental conditions, were executed (e.g., in parallel, sequentially, or mixed) &  \\ \hline
\textbf{Execution timeframe} & 5 & State over what timeframe the experiment was executed &  \\ \hline
\textbf{Pilots} & 6 & Describe if pilot studies were performed before the main experiment &  \\ \hline
\textbf{Returning workers} & 7 & Report the strategies used to prevent returning workers, i.e., workers who finish the experiment and then reenter it later because the study was still running &  \\ \hline
\multicolumn{4}{|l|}{} \\ \hline
\rowcolor[HTML]{CFE2F3} 
\multicolumn{4}{|c|}{\cellcolor[HTML]{CFE2F3}\textbf{Crowd}} \\ \hline
\textbf{Target population} & 8 & Describe the criteria used to determine the workers who are allowed to participate (e.g., acceptance rate, tasks completed, demographics, working environment). And also indicate
the strategy used to identify such workers &  \\ \hline
\textbf{Sampling mechanism} & 9 & Report what strategies were used to recruit a diverse or representative set of workers from the target population &  \\ \hline
\multicolumn{4}{|l|}{} \\ \hline
\rowcolor[HTML]{CFE2F3} 
\multicolumn{4}{|c|}{\cellcolor[HTML]{CFE2F3}\textbf{Task}} \\ \hline
\textbf{Task interface} & 10 & Report and show the task interface as seen by workers &  \\ \hline
\textbf{Task interface source} & 11 & Provide a link to an online repository with the source code of the task interface (typically a combination of HTML, CSS, and JavaScript) &  \\ \hline
\textbf{Instructions} & 12 & Describe and show the instructions of the task as seen by workers &  \\ \hline
\textbf{Reward strategy} & 13 & State the mechanisms used to reward and motivate workers (e.g., payments) &  \\ \hline
\textbf{Time allotted} & 14 & Report if a time constraint was defined for workers to complete the task (if so, describe also how much) &  \\ \hline
\multicolumn{4}{|l|}{} \\ \hline
\rowcolor[HTML]{CFE2F3} 
\multicolumn{4}{|c|}{\cellcolor[HTML]{CFE2F3}\textbf{Quality control}} \\ \hline
\textbf{Rejection criteria} & 15 & State the criteria used to accept or reject a contribution from a worker (e.g., workers can be allowed to submit the task and reject it afterward, submissions can be blocked based on prior rejections or on time spent on the task) &  \\ \hline
\textbf{Number of votes per item} & 16 & Describe, if applicable, how many workers solved the same item or data unit &  \\ \hline
\textbf{Aggregation method} & 17 & Report, if applicable, how the contributions from workers were aggregated (e.g., majority voting) &  \\ \hline
\textbf{Training} & 18 & State if workers performed a training session or pre-task qualification test. If so, describe 1) the training, 2) the items used as the training set, and 3) if it was performed before or as part of the task &  \\ \hline
\textbf{In-task checks} & 19 & Report the mechanisms embedded in the task to guard the quality of the results. Also, state if and how workers were allowed to revise their answers. In case gold items or attention checks were used, describe how these items were selected, how frequently they appear, and the threshold used to filter out workers underperforming on these items. &  \\ \hline


\textbf{Post-task checks} & 20 & Report the steps performed upon task completion to safeguard the quality of the results (e.g., post hoc analysis) &  \\ \hline
\textbf{Dropouts prevention mechanisms} & 21 & Indicate the strategies used to deal with worker dropouts (i.e., workers who leave the task unfinished) &  \\ \hline
\multicolumn{4}{|l|}{} \\ \hline
\rowcolor[HTML]{CFE2F3} 
\multicolumn{4}{|c|}{\cellcolor[HTML]{CFE2F3}\textbf{Outcome}} \\ \hline
\textbf{Number of participants} & 22 & Indicate how many workers participated in the experiment (in total and per condition) &  \\ \hline
\textbf{Number of contributions} & 23 & Report the number of contributions (e.g., votes) in total and per condition &  \\ \hline
\textbf{Excluded participants} & 24 & Indicate the number of participants Nt considered for the data analysis, including the reason for exclusion. &  \\ \hline
\textbf{Discarded data} & 25 & State the number of contributions excluded before the data analysis &  \\ \hline
\textbf{Dropout rate} & 26 & Describe the dropout rate of the participants in the experimental conditions. If applicable, also show breakdowns per milestone of progress within the task (e.g., after 2, 3, and 5 questions). &  \\ \hline
\textbf{Participant demographics} & 27 & Report the demographics of the participants (e.g., age, country, language) &  \\ \hline
\textbf{Data processing} & 28 & Report any data transformation, augmentation, and/or filtering step performed on the raw dataset obtained from the crowdsourcing platform. &  \\ \hline
\textbf{Output dataset} & 29 & Provide a link to the dataset resulting from the experiment. Also indicate if the dataset contains the aggregated or individual contributions from workers &  \\ \hline
\multicolumn{4}{|l|}{} \\ \hline
\rowcolor[HTML]{CFE2F3} 
\multicolumn{4}{|c|}{\cellcolor[HTML]{CFE2F3}\textbf{Requester}} \\ \hline
\textbf{Platform(s) used} & 30 & Indicate the crowdsourcing platform(s) selected for the experiment &  \\ \hline
\textbf{Implemented features} & 31 & Report any additional feature implemented to support the experiment, covering missing functionality from the selected platform(s) &  \\ \hline
\textbf{Fair compensation} & 32 & State whether workers were compensated fairly and according to legal minimum wage &  \\ \hline
\textbf{Requester-Worker interactions} & 33 & Describe concrete requester-worker interactions taking place as part of the experiment &  \\ \hline
\textbf{Privacy \& Data Treatment} & 34 & Report any relevant privacy regulations and methods used to comply, especially if the output is put online (e.g., the data could be aNnymized to meet privacy policies). &  \\ \hline
\textbf{Informed consent} & 35 & Indicate if an informed consent was used &  \\ \hline
\textbf{Participation awareness} & 36 & State if workers were informed they took part in an experiment &  \\ \hline
\textbf{Ethical approvals} & 37 & Report if the study received ethical approval from the corresponding institutional authority &  \\ \hline
\end{longtable}

\section{Identifying papers reporting crowdsourcing experiments} \label{sec:query-screening}


This section introduces the query that was used to retrieve papers (potentially) reporting crowdsourcing experiments and the exact eligibility criteria used to filter out retrieved articles.

\subsection{Scopus query}

\begin{Verbatim}[frame=single]
TITLE-ABS-KEY(crowdsource OR crowd-source OR crowd-sourcing OR
              crowdsourcing OR "human computation" OR
              crowdsourc* OR crowd-sourc* OR m*cro-task OR m*crotask)
AND
TITLE-ABS-KEY(experiment OR "experimental design" OR stud* OR evaluation OR
              intervention OR analysis)
AND
TITLE-ABS-KEY(user* OR behavio* OR worker*)
AND
(CONF(CHI) OR CONF(HCI) OR CONF(CSCW) OR CONF(WWW) OR CONF(HCOMP) OR
CONF(WSDM) OR CONF(CIKM) OR CONF(SIGIR) OR CONF(ICWE) OR CONF(IUI) OR
CONF(UIST) OR CONF(ICWSM) OR CONF("Human Factors"))
AND
(LIMIT-TO (DOCTYPE ,  "cp"))
\end{Verbatim}

\subsection{Screening instructions}

\begin{figure}[h]
  \centering
  \includegraphics[width=0.8\linewidth]{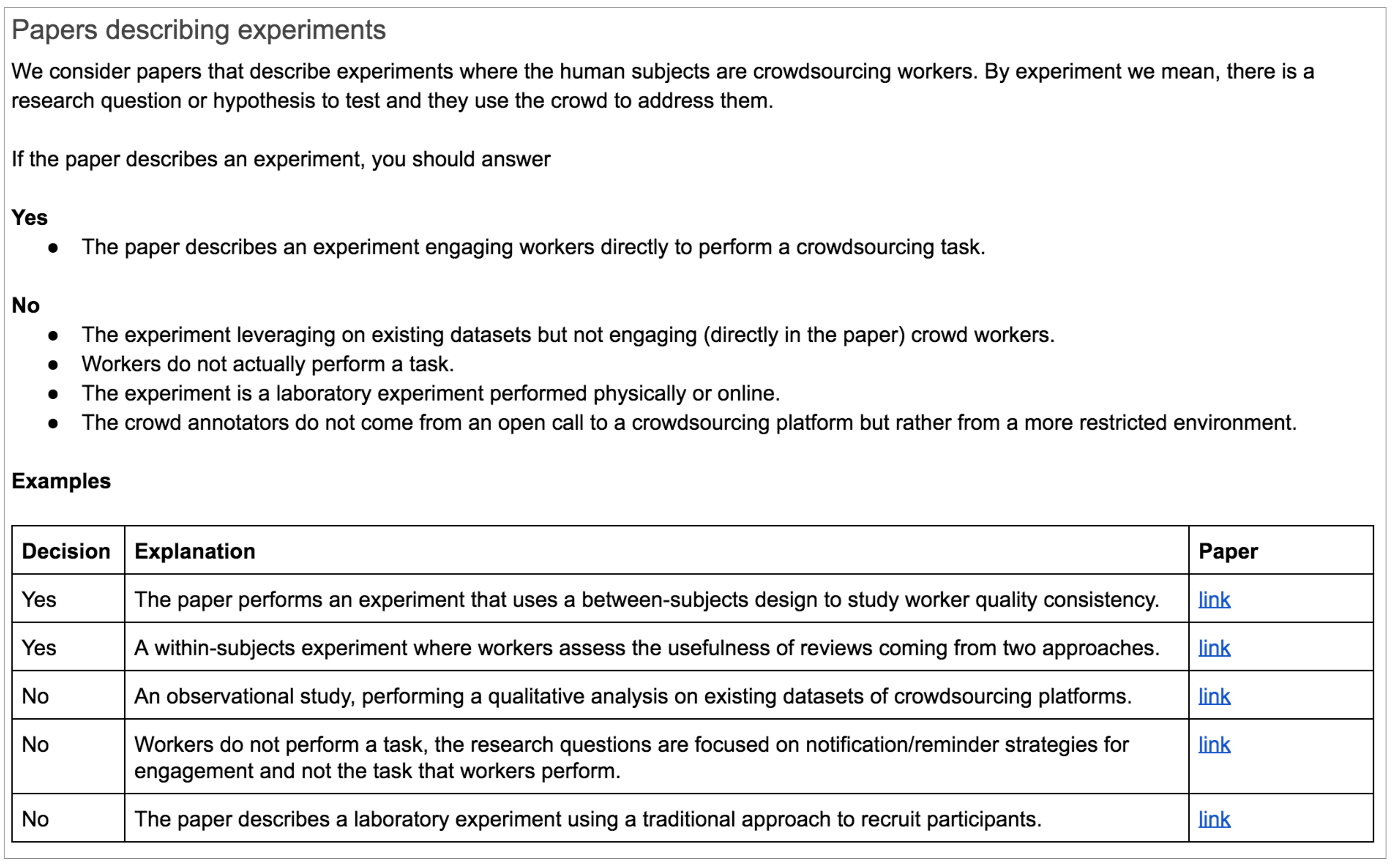}
  \caption{The screening instructions for identifying papers reporting crowdsourcing experiments.}
  \Description{The screening instructions for identifying papers reporting crowdsourcing experiments.}
  \label{fig:screening-instructions}
\end{figure}

Figure \ref{fig:screening-instructions} depicts the instructions used by the researchers to identify papers reporting crowdsourcing experiments.

\section{Interview protocol} \label{sec:interview-protocol}

The interview protocol can be found at \url{https://tinyurl.com/ReportingInterviewProtocol}.

\section{Applicability of attributes} \label{sec:applicability-attr}

\hl{
As we mentioned in Section \ref{sec:methods-state}, we considered 13 of the 39 attributes as potentially not applicable (N/A) based on the experiment's goal. Here, we detail these attributes.}

\hl{
The \textit{input dataset} was N/A if the paper does not necessarily use an input dataset for the crowdsourcing tasks. For example, in creative tasks, workers are just given instructions, and they provide input. The \textit{returning workers} attribute was N/A if the paper studied mechanisms to deal with workers that return to the experiment, or the study needed returning workers as part of their setup. The attributes in the quality control dimension were considered N/A if the paper actually studied quality control in crowdsourcing, including also strategies to deal with workers dropouts. The \textit{aggregation method} was N/A if the aggregation of contributions was not suitable for the experiment, and likewise, the \textit{gold items configuration} was N/A if the experiment did not use gold items.
The \textit{number of contributions} and \textit{discarded data} were N/A if they were just equal to the number of participants and excluded participants, respectively. An example of this is a study on worker behavior, which could collect a single contribution from each participant.
Finally, the \textit{data processing} was N/A if contributions were used as-is.
}




\end{document}